\documentclass[%
 aip,
 amsmath,amssymb,
 reprint,%
]{revtex4-1}

\usepackage[version=3]{mhchem}  % Formula subscripts using \ce{}
\usepackage{textcomp, gensymb}            % Add symbols like degree
\usepackage{xcolor}
\usepackage{physics}

\usepackage{graphicx}% Include figure files
\usepackage{dcolumn}% Align table columns on decimal point
\usepackage{bm}% bold math
\usepackage[utf8]{inputenc}
\usepackage[T1]{fontenc}
\usepackage{mathptmx}
\usepackage{etoolbox}

\makeatletter
\def\@email#1#2{%
 \endgroup
 \patchcmd{\titleblock@produce}
  {\frontmatter@RRAPformat}
  {\frontmatter@RRAPformat{\produce@RRAP{*#1\href{mailto:#2}{#2}}}\frontmatter@RRAPformat}
  {}{}
}%
\makeatother

\begin{document}

%\preprint{}

\title{Tuning Structural and Electronic Properties of Metal-Organic Framework 5 by Metal Substitution and Linker Functionalization} %Title of paper



\author{Joshua Edzards}
\email{joshua.edzards@uni-oldenburg.de}
\affiliation{Carl von Ossietzky Universit\"at Oldenburg, Institute of Physics, 26129 Oldenburg, Germany}
\author{Holger-Dietrich Sa{\ss}nick}
\affiliation{Carl von Ossietzky Universit\"at Oldenburg, Institute of Physics, 26129 Oldenburg, Germany}
\author{Julia Santana Andreo}
\affiliation{Carl von Ossietzky Universit\"at Oldenburg, Institute of Physics, 26129 Oldenburg, Germany}
\author{Caterina Cocchi}
\email{caterina.cocchi@uni-oldenburg.de}
\affiliation{Carl von Ossietzky Universit\"at Oldenburg, Institute of Physics, 26129 Oldenburg, Germany}


\date{\today}

\begin{abstract}
The chemical flexibility of metal-organic frameworks (MOFs) offers an ideal platform to tune structure and composition for specific applications, from gas sensing to catalysis and from photoelectric conversion to energy storage. This variability gives rise to a large configurational space that can be efficiently explored using high-throughput computational methods. In this work, we investigate from first principles the structural and electronic properties of MOF-5 variants obtained by replacing Zn with Be, Mg, Cd, Ca, Sr, and Ba, and by functionalizing the originally H-passivated linkers with CH$_3$, NO$_2$, Cl, Br, NH$_2$, OH, and COOH groups. To build and analyze the resulting 56 structures, we employ density-functional theory calculations embedded in an in-house developed library for automatized calculations. Our findings reveal that structural properties are mainly defined by metal atoms and large functional groups which distort the lattice and modify coordination. Stability is largely influenced by functionalization and enhanced by COOH and OH groups which promote the formation of hydrogen bonds. The charge distribution within the linker is especially influenced by functional groups with electron-withdrawing character while the metal nodes play a minor role. Likewise, the band-gap size is crucially determined by ligand functionalization. The smallest gaps are found with NH$_2$ and OH groups which introduce localized orbitals at the top of the valence band. This characteristic makes these functionalizations particularly promising for the design of MOF-5 variants with enhanced gas uptake and sensing properties.
\end{abstract}

\maketitle
\section{Introduction}
Metal-organic frameworks (MOFs) are a novel class of materials regarded with particular attention thanks to their unique structural and chemical properties.~\cite{furu+13sci} Microscopically, they are formed by a crystalline arrangement of metal atoms bound together by molecular linkers giving rise to a porous lattice.~\cite{jame+03csr} This peculiar characteristic is crucial in many technological fields, ranging from gas storage and sensing~\cite{rosi+03sci,koo+19chem,nguy+20ass} to photocatalysis~\cite{ma+21jacs,wu+20ccr} and from photoelectric conversion~\cite{prat+20jpcc,liu+23nl} to supercapacitors and batteries.~\cite{liu+10car,wang+16ccr} This variety of applications can be targeted by the large tunability of MOFs. The pore size is one of the main parameters that can be varied according to specific requirements. Its modulation can be achieved by tuning the lattice~\cite{yuan+10jacs,chen+16acie} or by modifying the organic ligands, for example, in terms of chain length,~\cite{lian+14cec} by altering their coordination,~\cite{du+13jacs,asse+15acie} or via substitution or functionalization.~\cite{zhao+11acr,zhan+17acie}

Among the numerous classes of MOFs that have been realized so far, MOF-5 is one of those that received the most attention. Since its first synthesis by solvothermal methods,~\cite{Li+99nat} it has shown to be particularly flexible and prone to functionalization thanks to its structure formed by Zn$_4$O clusters held together by 1,4-benzodicarboxyl acid.~\cite{kaye+07jacs} These properties turned out to be particularly favorable for gas uptake and storage.~\cite{li+09ijhe,yang+12ijhe,zhao+13iecr} Furthermore, it was shown that doping MOF-5 with transition-metal atoms triggers photoluminescence,~\cite{liu+11jmc,kato+18vac} thus promoting its application for sensing and other optically driven functionalities.~\cite{bota+10lan,yang+14jssc} 

The remarkable number of MOF-5 variants achieved by synthetic chemistry (see Ref.~\citenum{gang+22rscadv} for review) suggests that further tunability can be reached by means of computational screening. The automatization of \textit{ab initio} calculations based on density-functional theory (DFT) has paved the way for exploring materials on a quantum-mechanical level without the need for experimental parameters. This enables investigating chemical and structural variations in a predictive way to guide synthesis and laboratory characterization. This approach has been successfully applied to identify the most suitable MOF compositions for target applications out of a large configurational space~\cite{cane+13jmca,rose+19jcc} or even from a few tens of chemical variants.~\cite{sass+24ic} In both flavors, these studies have shown the potential of atomistic computational approaches to screen and characterize MOFs. 

In this work, we adopt automatized first-principles methods based on DFT to investigate the structural and electronic properties of MOF-5 upon metal-node exchange and ligand functionalization. 
Taking advantage of an in-house developed computational infrastructure for high-throughput screening and automated analysis,~\cite{Sassnick2022} we explore 56 different configurations obtained by substituting Zn with Be, Mg, Cd, Ca, Sr, and Ba, and by functionalizing the organic linker with \ce{CH3}, \ce{NO2}, Cl, Br, \ce{NH2}, OH, and COOH, in addition to native H.
We analyze the effects of these chemical modifications on the structural properties of the compounds, focusing in particular on the variation of the lattice constant and the rearrangement of the functionalized ligands in the scaffold, which in turn can affect metal coordination.
Next, we examine the stability of the predicted systems in terms of formation energy per atom and, finally, we inspect the electronic properties of the considered MOF-5 variants. Partial charge analysis gives insight into the nature of the chemical bonds while the projected density of states provides information about the size of the band gaps and the character of the electronic states.
This analysis offers valuable indications to identify materials with optimal characteristics for gas storage and sensing. 

%%%%%%%%%%%%%%%%%%%%%%%%%%%%%%%%%%%%%%%%%%%%%%%%%%%%%%%%%%%%%%%%%%%%%
%% Systems and Methods
%%%%%%%%%%%%%%%%%%%%%%%%%%%%%%%%%%%%%%%%%%%%%%%%%%%%%%%%%%%%%%%%%%%%%
\section{Methods}
\label{sec:systems-methos}

%%%%%%%%%%%%%%%%%%%%%%%%%%%%%%%%%%%%%%%%%%%%%%%%%%%%%%%%%%%%%%%%%%%%%
\subsection{Construction of the Systems}

\begin{figure*}
    \centering
    \includegraphics[width=\textwidth]{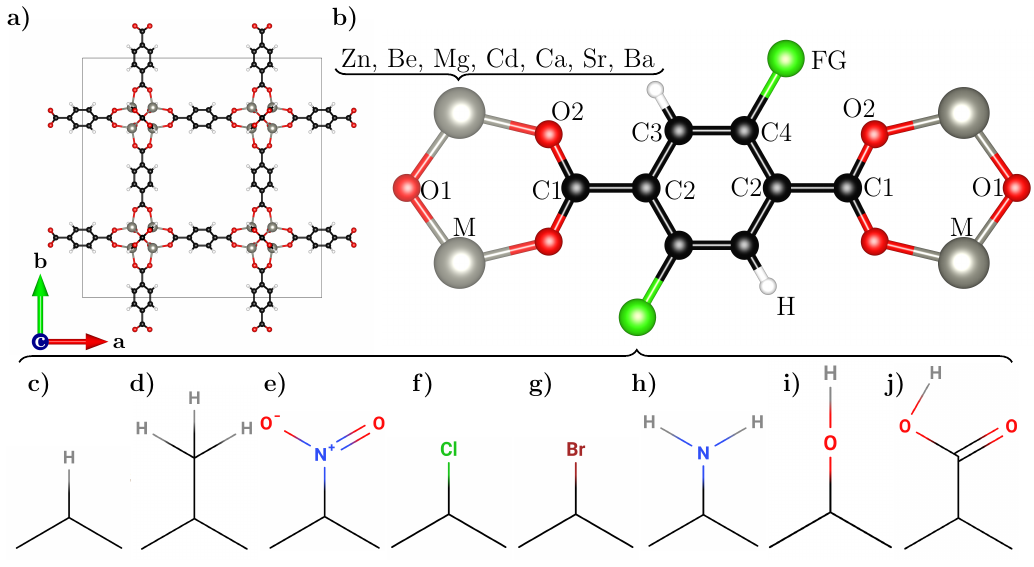}
    \caption{a) Ball-and-stick representation of the cubic conventional cell of MOF-5. b) Organic BDC linker molecule with diagonal functionalization. Carbon atoms are depicted in black, hydrogen atoms in white, metal atoms in gray (Zn in MOF-5, Be, Mg, Cd, Ca, Sr, and Ba), and the functional groups (FG) in green: c) H, d) \ce{CH3}, e) \ce{NO2}, f) Cl, g) Br, h) \ce{NH2}, i) OH, and j) COOH. The label of each element is dedicated to the symmetry of elements considered equal coordination. Graphics created with VESTA.~\cite{Momma2011}}
    \label{fig:MOF_5_structure}
\end{figure*}

As a starting point for the construction of MOF-5 and its derivatives considered in this work, we take the structure proposed by Butler \textit{et al}.~\cite{Butler2014, Butler2020}
It is a face-centered cubic (FCC) crystal belonging to the $Fm\overline{3}m$ space group (see Fig.~\ref{fig:MOF_5_structure}a) with experimental lattice constant $a=25.87$~\AA{}.~\cite{Li+99nat}
The Zn atoms are connected through the linker molecules 1,4-benzene-dicarboxylate (BDC).
The functionalization sites are marked in green in Fig.~\ref{fig:MOF_5_structure}b.
To explore metal node exchange, we replace Zn with the isoelectronic elements Be, Mg, Cd, Ca, Sr, Ba (see Fig.~\ref{fig:MOF_5_structure}b).
Given the different atomic radii of the aforementioned species,~\cite{Cordero2008} the atomic positions and the lattice parameters are adjusted with the aid of an in-house developed \texttt{python} library.~\cite{Sassnick2022}
To study the effects of ligand substitutions, we replace two H atoms in the phenyl ring with functional groups including \ce{CH3}, \ce{NO2}, Cl, Br, \ce{NH2}, OH, and COOH, depicted in Figures~\ref{fig:MOF_5_structure}d-j, respectively.
In the construction of these systems, the initial bond lengths between the carbon atom anchoring the functional group (C4) and the group itself are set to 1.00~\AA{} with H (C-H bond), 1.53~\AA{} with \ce{CH3} and COOH (C-C bond), 1.93~\AA{} with Br (C-Br bond), 1.76~\AA{} with Cl (C-Cl bond), 1.47~\AA{} with \ce{NH2} and \ce{NO2} (C-N bond), and 1.41~\AA{} with OH (C-O bond).~\cite{Haynes2014}

%%%%%%%%%%%%%%%%%%%%%%%%%%%%%%%%%%%%%%%%%%%%%%%%%%%%%%%%%%%%%%%%%%%%%
\subsection{Computational Details}

All DFT calculations presented in this work were performed using \texttt{CP2K}~\cite{Kuehne2020} version 9.1 with Goedecker-Teter-Hutter pseudopotentials~\cite{Goedecker1996} and the Perdew-Burke-Ernzerhof (PBE) functional~\cite{perd+96prl} augmented with the Grimme's D3 scheme to account for dispersive interactions.~\cite{Grimme2010}
The MOLOPT triple-$\zeta$ basis set with double polarization~\cite{Jensen2007} is used in all runs except for the pre-optimization step where the double-$\zeta$ set with single polarization is adopted. 
The cutoff values for the plane wave and the relative are set to 600 Ry and 100 Ry, respectively.
A k-mesh of equidistant points separated by 0.1~\AA$^{-1}$ is constructed using (2$\times$2$\times$2) k-points within the Monkhorst-Pack scheme~\cite{Monkhorst1976}.
The parameters for the pre-optimization step are decreased to 300~Ry for the cutoff and 0.4~\AA$^{-1}$ for the k-point separation; the thresholds for pressure and force are set to 200~bar and 0.005~Hartree/bohr, respectively.
In the volume optimization, the latter values are decreased to 100~bar (pressure) and 0.00048~Hartree/bohr (force), and the angles are kept fixed to preserve the symmetry of the input structure.
To calculate the pDOS, a supercell of size (2$\times$2$\times$2) is created to mimic the k-point sampling.
Partial charges are computed with the software \texttt{Critic2}~\cite{otero2014}, version 1.1, using the Bader scheme~\cite{Bader92cpl} and the Yu-Trinkle integration method.~\cite{Yu2011}

%%%%%%%%%%%%%%%%%%%%%%%%%%%%%%%%%%%%%%%%%%%%%%%%%%%%%%%%%%%%%%%%%%%%%
%% Results and Discussion
%%%%%%%%%%%%%%%%%%%%%%%%%%%%%%%%%%%%%%%%%%%%%%%%%%%%%%%%%%%%%%%%%%%%%
\section{Results and Discussion}
\label{sec:results}
%%%%%%%%%%%%%%%%%%%%%%%%%%%%%%%%%%%%%%%%%%%%%%%%%%%%%%%%%%%%%%%%%%%%%
\subsection{Structural Optimization}
\label{ssec:struct}

\begin{figure*}
    \centering
    \includegraphics[width=\textwidth]{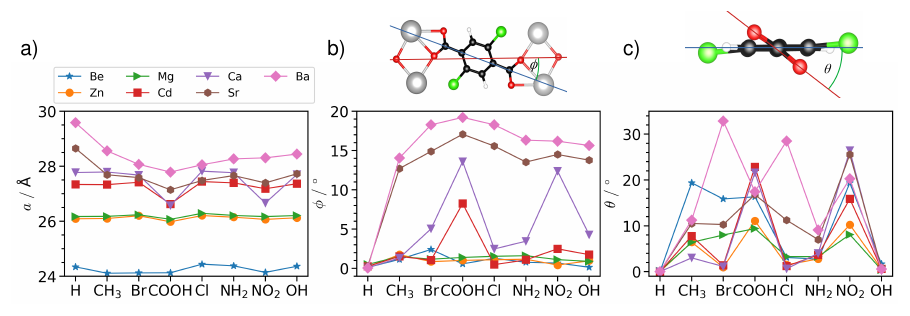}
    \caption{a) Lattice constant $a$ of the optimized cubic conventional cell of MOF-5-like structures at varying metal nodes and linker functionalization; b) Average rotation angle $\phi$ around the centered normal of BDC; c) rotation angle $\theta$ between the oxygen atoms and the aryl group in the BDC backbone.}
    \label{fig:struct}
\end{figure*}

%%%%%%%%%%%%%%%%%%%%%%%%%%%%%%%%%%%%%%%%%%%%%%%%%%%%%%%%%%%%%%%%%%%%%
%% Lattice Constant
We start our analysis by investigating how the structural properties of MOF-5 are influenced by metal node exchange and linker functionalization.
%Since the van der Waals radii of each metal node differ we can expect a change in the lattice constant and further investigate the impact of the ligand substitution. 
To do so, we examine the lattice parameter $a$ of the cubic crystal shared by all the considered structures (see Fig.~\ref{fig:struct}a and Table S1 in the Supporting Information for the raw data).
The unit cell size is primarily determined by the van der Waals radius of the metal node while the termination of the molecular ligands plays a secondary role.
Hosting the smallest metallic species among the considered ones, Be-based MOFs exhibit the smallest lattice constant $a$, ranging between 24 and 24.5~\AA{} depending on the linker functionalization. 
Structures with Mg and Zn nodes exhibit values of $a$ around 26~\AA{} with fluctuations induced by molecular functionalization of the order of 0.2~\AA{}.
As expected, the inclusion of Ba leads to the largest unit cells ($a$ ranging between 20 and 30~\AA{}) while scaffolds embedding Cd, Ca, and Sr present intermediate values of $a$ comprised between 26.5~\AA{} and 28.7~\AA{}.
In the systems with larger volumes, the influence of linker substituents is more pronounced. 

To better understand these characteristics, we inspect in more detail the predicted structures and specifically the angle formed by the ligands with the centered normal axis of the BDC molecule $\phi$ (see Fig.~\ref{fig:struct}b) and the rotation $\theta$ between the oxygen atoms and the plane of the aryl group (Fig.~\ref{fig:struct}c).
In all H-terminated structures, both angles are zero meaning that the linker molecule is unaltered compared to its initial configuration regardless of the size of the metal node.
Ligand functionalization even with the atomic species Cl and Br distorts the BDC: these structural variations are closely related to the van der Waals radius of the metal ions and with the characteristics of the functional group. 
Focusing on MOF-5 with Sr and Ba, we find consistent angles $\phi$ between 12$^{\circ}$ and 20$^{\circ}$ for all terminations (Fig.~\ref{fig:struct}b).
We recall that with these two compositions, the lattice constant decreases substituting H with any other considered atom or group (see Fig.~\ref{fig:struct}a).
These trends are explained by the increased coordination of these large metallic nodes~\cite{Shannon1976} (see sketch in Fig.~\ref{fig:struct}c), which distorts the ligand with respect to its ideal configuration in MOF-5 while concomitantly reducing the unit cell volume. 
The microscopic origin of this sizeable structural modification can be traced back to the charge unbalance between the overall positively charged metal nodes and the negatively charged functional groups. 
We will come back to this point later in the analysis of the partial charge distribution.

These effects are much less pronounced with smaller metallic nodes leaving the pores big enough to host the functional groups without significant structural variations.
Exceptions are found for the COOH termination due to its particularly large size, with rotations $\phi=8^{\circ}$ and $\phi=13^{\circ}$ appearing in the Cd- and Ca-based MOF-5, respectively (see Fig.~\ref{fig:struct}b).
\ce{NO2} functionalization induces an equally large distortion in the Ca-based structure: in this case, the strong electronegativity of the termination is responsible for changing the coordination of the metallic species, and, consequently, for the linker rotation.

The oscillations of the angle $\theta$ (Fig.~\ref{fig:struct}c) are mainly associated with the steric and electrostatic hindrance between the functional groups and the oxygen atoms embedded in the BDC backbone. 
For this specific characteristic, a clear trend related to the size of the metal node is not immediately identifiable.
The largest values of $\theta$ are generally obtained with COOH and \ce{NO2} functionalizations and their chemical-physical origin can be traced back to the size of the former group and the stark electronegativity of both. 
Sizeable distortions close to $30^{\circ}$ are found in the Ba-based MOF-5 functionalized with Cl and Br: they can be ascribed to the combined effect of enhanced coordination of the metal node and the pronounced electron-withdrawing nature of the halogen atoms.
It is worth noting that with the same terminations, the structures embedding Ca, Cd, and Zn exhibit values of $\theta$ close to zero.
Undistorted molecular backbones ($\theta=0^{\circ}$) are obtained with H and OH regardless of the metal node, in line with the behavior of $\phi$ discussed above. 
Backbone rotations with $\theta<20^{\circ}$ and $\theta<10^{\circ}$ are induced by \ce{CH3} and \ce{NH2} functionalization, once again consistent with the size and the electronegativity of these groups.

Before closing this section, we briefly discuss our results in the context of the available literature.
Earlier DFT studies of H-terminated MOF-5 performed with the PBE functional~\cite{Yang2011, Yang2010, Srepusharawoot2008} reported lattice constants in agreement with those presented in Fig.~\ref{fig:struct}a.
Distorted BDC linkers with angles $\theta$ up to $90^{\circ}$ were found experimentally in \ce{CH3}-functionalized MOF-5~\cite{Rowsell2004} and theoretically predicted in halogen-terminated systems~\cite{Yang2016}.
These rotations are much more pronounced than those reported in Fig.~\ref{fig:struct}c, where, in contrast to the above-mentioned studies focused on fully functionalized linkers, the data are referred to BDC with only two substituents.

\subsection{Formation Energy}
\label{ssec:energetics}

We continue our analysis by focusing on the stability of the considered structures.
We evaluate this property with the formation energy per atom computed according to Eq.~\eqref{eq:Formation}: %in the ground state is subtracted from the total energy of the system.
\begin{equation}\label{eq:Formation}
    \begin{aligned}
        E_{form}&=E(2[\text{M}_4\text{O}(\text{BDC-FG})_3])
        -\frac{8}{94+12n}E(\text{M})\\
        &-\frac{26}{94+12n}E(\text{O})
        -\frac{48}{94+12n}E(\text{C})\\
        &-\frac{12}{94+12n}E(\text{H})
        -\frac{12}{94+12n}\sum_{i=1}^nE(\text{A}_i^{\text{FG}}),
    \end{aligned}
\end{equation}
where M denotes the metal species, BDC the organic linker, FG the functional group as a whole, and A the atoms therein; O, C, and H are the conventional chemical symbols for oxygen, carbon, and hydrogen, and the index $n$ labels the number of atoms of the given type.

\begin{figure}
    \centering
    \includegraphics[width=0.5\textwidth]{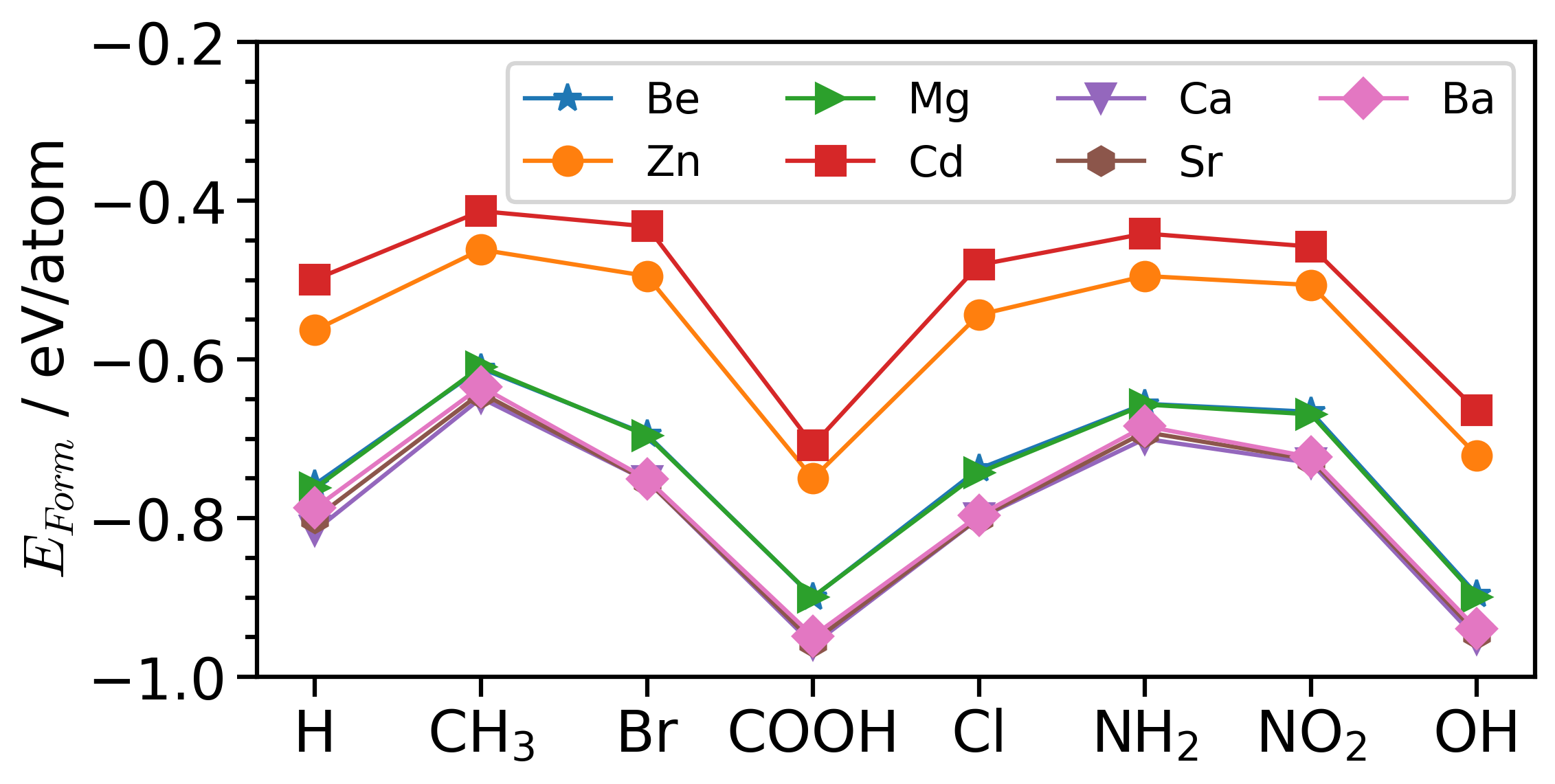}
    \caption{Formation energies of the considered MOF-5 derivatives.}
    \label{fig:Formation_Energy}
\end{figure}

The results reported in Fig.~\ref{fig:Formation_Energy} (raw data in Table S4), showing only negative values of $E_{form}$, indicate that all considered structures are stable regardless of the choice of the metal node or the functionalization.
Furthermore, some clear trends are visible.
The MOF-5 structures including alkali-earth metallic species are the most stable ones, exhibiting the same formation energy (differences for equally functionalized systems are up to 20~meV) for the same ligand termination.
In particular, we can identify almost identical values of $E_{form}$ obtained with Ca, Sr, and Ba, giving rise to the most stable structures, and with Be and Mg. 
The same qualitative trend with respect to the linker functionalization is found also with the other considered metal nodes, Zn and Cd, leading to structures with approximately half the stability of the other ones. 

The influence of the BDC terminations on the stability of the considered MOFs is remarkable, with variations of hundreds of meV for a given metal node (see Fig.~\ref{fig:Formation_Energy}).
The presence of COOH and OH functional groups gives rise to the most stable structures.
This behavior can be explained by the formation of hydrogen bonds, which are known to stabilize the MOFs~\cite{sass+24ic}.
\ce{NH2} and \ce{CH3} terminations promote hydrogen bonds too, but their strength is insufficient to increase the stability of the crystals.
Halogen terminations lead to MOFs with similar stability as the H-passivated ones. 
This finding may contradict the physicochemical intuition based on the strength of C-Br, C-Cl, and C-H bonds~\cite{blan+2003acr} but in interpreting the results shown in Fig.~\ref{fig:Formation_Energy}, one should consider that kinetic and thermodynamic contributions, which are vital for chemical reactions, are not included in these calculations.
\ce{NO2} functionalization leads to structures that are energetically close to those predicted with \ce{NH2} but less stable than those with the reference H-termination: this behavior can be understood by considering the electrostatic repulsion induced by such a highly electronegative group.

%In general, all metal nodes represent negative formation energies $E_F$ and thus are stable for synthesis. The results are plotted in Fig.~\ref{fig:Formation_Energy} and the raw data is listed in SI-Table~\ref{SI_tab:Formation_Energy} The earth alkali metals form a group and have higher negative formation energy $E_F$ than the metal nodes of group 12. This indicates a high stability of alkaline earth metals. But each metal node follows similar trends for the functional groups and thus only the trend of the functional groups are further discussed. Additionally, the functional groups have a significant impact on the stability. The contribution to the lowest formation energy $E_F$ comes from the carboxylic group and the hydroxyl group since they form hydrogen bonds with the first oxygen of the BDC-linker molecule (O2) which then contributes to the stability of the systems. Again, the amino groups form hydrogen bonds but the distances compared to the other two functional groups are larger. This leads to negligible contributions to the stability and is more or less equally stable like the methyl group and nitrogen dioxide, which are the least stable functionalization. Hydrogen, bromide, and chlorine donate comparably to the stability. The only difference appears due to the different binding energies between the C-Br, C-Cl, and C-H bonds. During the calculation, the temperature and the reaction kinetics, that will impact the reaction during the synthesis, are not considered in this work. 

%%%%%%%%%%%%%%%%%%%%%%%%%%%%%%%%%%%%%%%%%%%%%%%%%%%%%%%%%%%%%%%%%%%%%
\subsection{Partial Charge Analysis}
\label{ssec:charges}

\begin{figure}
    \centering
    \includegraphics[width=0.5\textwidth]{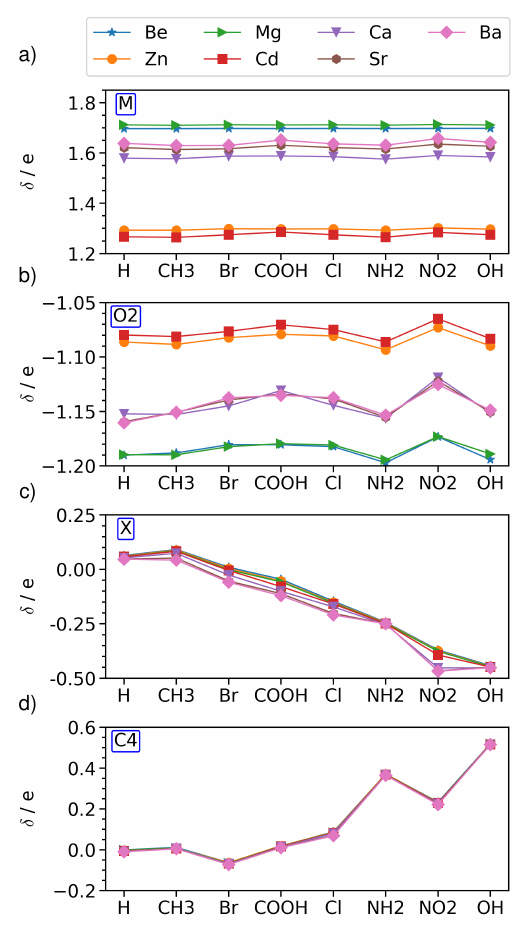}
    \caption{Partial charges $\delta$ on a) the metal species (M), b) the oxygen atom (O2) in the BDC backbone, c) the whole functional group (X), and d) the carbon atom (C4) in BDC computed for all the considered MOF-5 derivatives. 
    }
    \label{fig:partial_charges_X_C4_M_O2}
\end{figure}

Equipped with the knowledge of the structural properties, we proceed with the partial-charge analysis.
This way, we aim to gain a better understanding of the electron density distribution.
The adopted Bader scheme~\cite{Bader92cpl} is particularly suited to deal with periodic systems and is known to provide reliable results~\cite{edza+2023jpcc, sass+24ic}.
With this method, we evaluate the relative amount of electronic charge on the atomic species of MOF-5 and its related compounds with ligand substitution. 
The results reported in Fig.~\ref{fig:partial_charges_X_C4_M_O2} (raw data in SI-Tables S10 (M), S12 (O2), S13 (X) and S8 (C4)) correspond to the partial charges averaged on all atoms of a given species within the unit cell adopted in the simulations.

%%%%%%%%%%%%%%%%%%%%%%%%%%%%%%%%%%%%%%%%%%%%%%%%%%%%%%%%%%%%%%%%%%%%%
%% Metal
The metal nodes embedded in all structures exhibit a large positive charge, ranging from approximately 1.3~e$^-$ in Cd and Zn to values of about 1.7~e$^-$ for the alkali-earth species.
This distribution in two groups is consistent with the electronic configuration of the atoms. 
The charge of the transition metals Cd and Zn is screened by their full $d$-shell and it is thus lower compared to that of the alkali-earth species.
The functional groups do not significantly affect the partial charges at the metal nodes.
They merely induce oscillations in their values with maxima achieved in the presence of COOH and \ce{NO2} groups due to their influence on the metallic coordination in the MOF.
This result contributes to explaining the structural trends discussed above.

%Starting with the metal node shown in Figure \ref{fig:partial_charges_X_C4_M_O2}a), each metal node is distinct and partially positively charged starting with the lowest charged cadmium and ending with the highest charged magnesium. We can differentiate the metal nodes into two groups. The first group forms the earth alkaline metals and the second group is based on the metals from group 12.\textcolor{red}{In each group, the charge follows the trend of electronegativity but also the distance to the nearest neighboring oxygen plays a huge role. Thus beryllium and magnesium are the most positively charged and barium, strontium, and calcium are the least positively charged in this group.Cadmium and zinc follow the same trend but the electrons are bonded strongly due to the additionally filled d-orbital, the contribution to the nearest oxygen is smaller.} The functional groups seem to not affect the partial charges at the metal nodes. Since the functional groups are either too far away to interact with the metal nodes or their charge is too weak it is not noticeable by the metal node. An increase of positive partial charge is noticeable for the carboxylic group and nitrogen dioxide for several metal nodes. Since the functional groups COOH and \ce{NO2} bind with the metal node, an increase in charge is expected. The metal node contributes some electron charge to the functional group.

%%%%%%%%%%%%%%%%%%%%%%%%%%%%%%%%%%%%%%%%%%%%%%%%%%%%%%%%%%%%%%%%%%%%%
%% Oxygen (O2)
The negative values of the partial charges of the oxygen atoms in the BDC, ranging between -1.20~e$^-$ and -1.05~e$^-$, indicate the predominant ionic nature of the metal-linker bonds in MOF-5 and its derivatives (see Fig.~\ref{fig:partial_charges_X_C4_M_O2}b).
Again, our findings can be grouped according to the metallic species: structures embedding Be and Mg, which are the smallest alkali-earth elements, exhibit the most negatively charged oxygens.
Results obtained in the Ca-, Sr-, and Ba-based structures are almost identical and form another group with partial charges $\delta$ fluctuating around -1.15~e$^-$ depending on the linker functionalization.
Finally, the group-12 elements Zn and Cd lead to less negative partial charges on the oxygen atoms of BDC, with values comprised between -1.10~e$^-$ and -1.05~e$^-$.
These trends mirror those discussed for the metallic species in Fig.~\ref{fig:partial_charges_X_C4_M_O2}a and confirm the dominant ionic nature of the metal-oxygen coordination.
The influence of the functional groups on the partial charges on oxygen is equivalent to that on the metal nodes, with maxima obtained with \ce{NO2}-terminations.

%The charge distribution from the metal node to the functional group is three orders of magnitude smaller than the contribution to the oxygen (O2) corresponding to the BDC-linker. Figure \ref{fig:partial_charges_X_C4_M_O2}b) shows the distribution of the partially charged oxygen (O2) depending on the functional group and the metal node. The oxygen has a huge impact on the metal node and leads to a partially positively charged metal node and leads to the opposite order of metal nodes compared to the charge in Figure \ref{fig:partial_charges_X_C4_M_O2}a). Again, the carboxylic group and nitrogen dioxide influence slightly the charge of the oxygen compared to the rest of the functional groups due to their interaction.

%%%%%%%%%%%%%%%%%%%%%%%%%%%%%%%%%%%%%%%%%%%%%%%%%%%%%%%%%%%%%%%%%%%%%
%% Functional Group (X)
The partial charges on the whole functional groups displayed in Fig.~\ref{fig:partial_charges_X_C4_M_O2}c  exhibit positive and negative values spanning the range between -0.5~e$^-$ and 0.1~e$^-$ according to their chemical nature.
The influence of the metal node is negligible with H, \ce{NH2}, and OH terminations while variations up to 100~meV are induced in the other systems and most pronounced in the presence of COOH and \ce{NO2} functionalizations.
This behavior can be related to the change in the metal coordination induced by these groups which is most pronounced with the largest metallic species: the closer proximity to these positively charged atoms (see Fig.~\ref{fig:partial_charges_X_C4_M_O2}a) enhances the negativity of their partial charges.
The results shown in Fig.~\ref{fig:partial_charges_X_C4_M_O2}c follow intuition, with positive charges around 0.05~e$^-$ on H atoms and not exceeding 0.1~e$^-$ on \ce{CH3}: both terminations are known for their electron-donating character.
All the other functionalizations exhibit negative partial charges with the sole exception of Br in the MOFs hosting the smallest metal nodes where $\delta \rightarrow 0$.
COOH and Cl are both slightly negatively charged in agreement with their mild electron-withdrawing nature.
The larger electronegativity of \ce{NO2} is reflected in the values of its partial charges fluctuating around -0.4~e$^-$ depending on the metal node.
Negative charges are found also on the electron-donating groups \ce{NH2} and OH.
This counter-intuitive finding can be explained by examining the results obtained for their neighboring carbon atoms C4 (see Fig.~\ref{fig:MOF_5_structure}b).

%Figure \ref{fig:partial_charges_X_C4_M_O2}c) represents the partial charges for the functional groups (X). The (X) sums over the average charge contribution of each atom and leads to a total partial charge of the functional group. The charge decreases with the choice of the functional groups and represents a wide spectrum of electron-donating and withdrawing behavior. Thus, hydrogen, the methyl group, and bromide are the least charged functional groups, and nitrogen dioxide and the hydrogen group contain the highest negative charge. Overall, the metal node has a small impact on the partial charges at the functional groups. Only the carboxylic group and nitrogen dioxide show different behavior based on their rotation and thus the interaction with the nearest oxygen (O2). This trend can be seen in calculated charges for the metal node and the corresponding oxygen (O2) in Figure \ref{fig:partial_charges_X_C4_M_O2}a) and b). 

%%%%%%%%%%%%%%%%%%%%%%%%%%%%%%%%%%%%%%%%%%%%%%%%%%%%%%%%%%%%%%%%%%%%%
%% Carbon (C2)

These species exhibit positive partial charges almost in all considered systems (see Fig.~\ref{fig:partial_charges_X_C4_M_O2}d), mirroring the positive charges with the same magnitude computed for these species (see Fig.~\ref{fig:partial_charges_X_C4_M_O2}c).
Indeed, the only negative values around -0.1~e$^-$ are obtained with Br while almost charges up to 0.1~e$^-$ are found on H-, \ce{CH3}-, COOH- and Cl-terminated C4 atoms.
Sizeable positive charges around 0.2, 0.4, and 0.6~e$^-$ are computed with \ce{NO2}, \ce{NH2}, and OH functionalizations.
In these results, the influence of the metal node is completely negligible indicating that the charge distribution within the carbon backbone of the BDC molecule is independent of this parameter and solely determined by the nearest chemical environment (namely, the functionalization) of the linker.
The complete charge analysis of BDC is reported in the SI (see Tables S5-S9, S11 and S12).

%%%%%%%%%%%%%%%%%%%%%%%%%%%%%%%%%%%%%%%%%%%%%%%%%%%%%%%%%%%%%%%%%%%%%%
\subsection{Electronic Properties}
\label{ssec:electronic}

\begin{figure}
    \centering
    \includegraphics[width=0.5\textwidth]{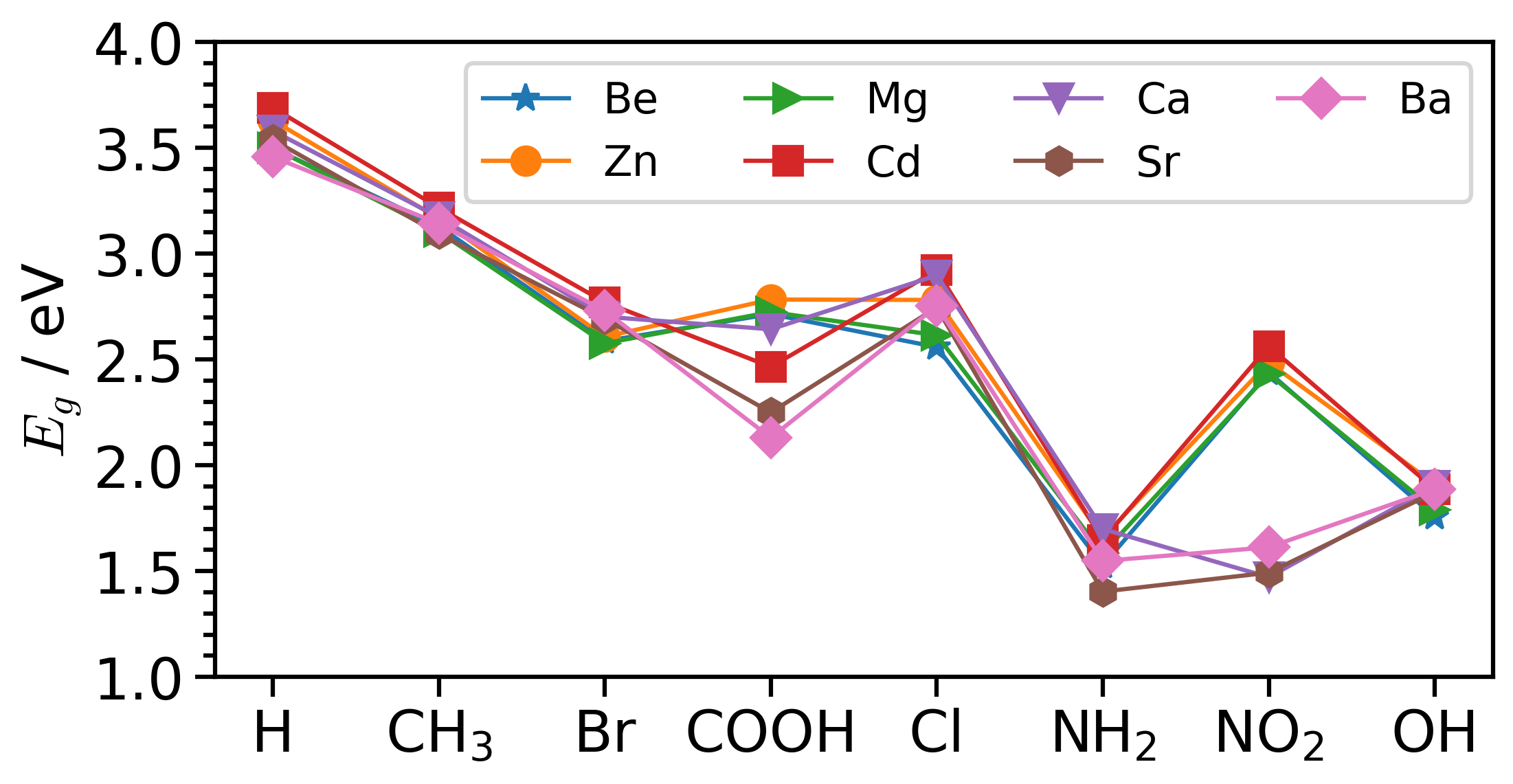}
    \caption{Band gaps ($E_{g}$) for MOF-5 and its derivatives calculated with DFT (PBE functional).}
    \label{fig:Band_gap}
\end{figure}

%%%%%%%%%%%%%%%%%%%%%%%%%%%%%%%%%%%%%%%%%%%%%%%%%%%%%%%%%%%%%%%%%%%%%
%% Bandgap
We conclude our investigation with the analysis of the electronic properties of MOF-5 and its derivatives starting from the band gaps (see Fig.~\ref{fig:Band_gap} and SI-Table S14).
It should be recalled that these values are obtained from DFT calculations using the PBE functional~\cite{perd+96prl} which is known to largely underestimate the band-gap size, on the order to 50\%, compared to the experimental value~\cite{Perdew1983}.
However, since it is expected that the same error affects each system, the trends can be considered reliable and interpreted accordingly.
%The calculated values are presented in Figure \ref{fig:Band_gap} and the raw data can be retrieved in SI-Table~\ref{SI_tab:Band_gab}.

\begin{figure*}
    \centering
    \includegraphics[width=\textwidth]{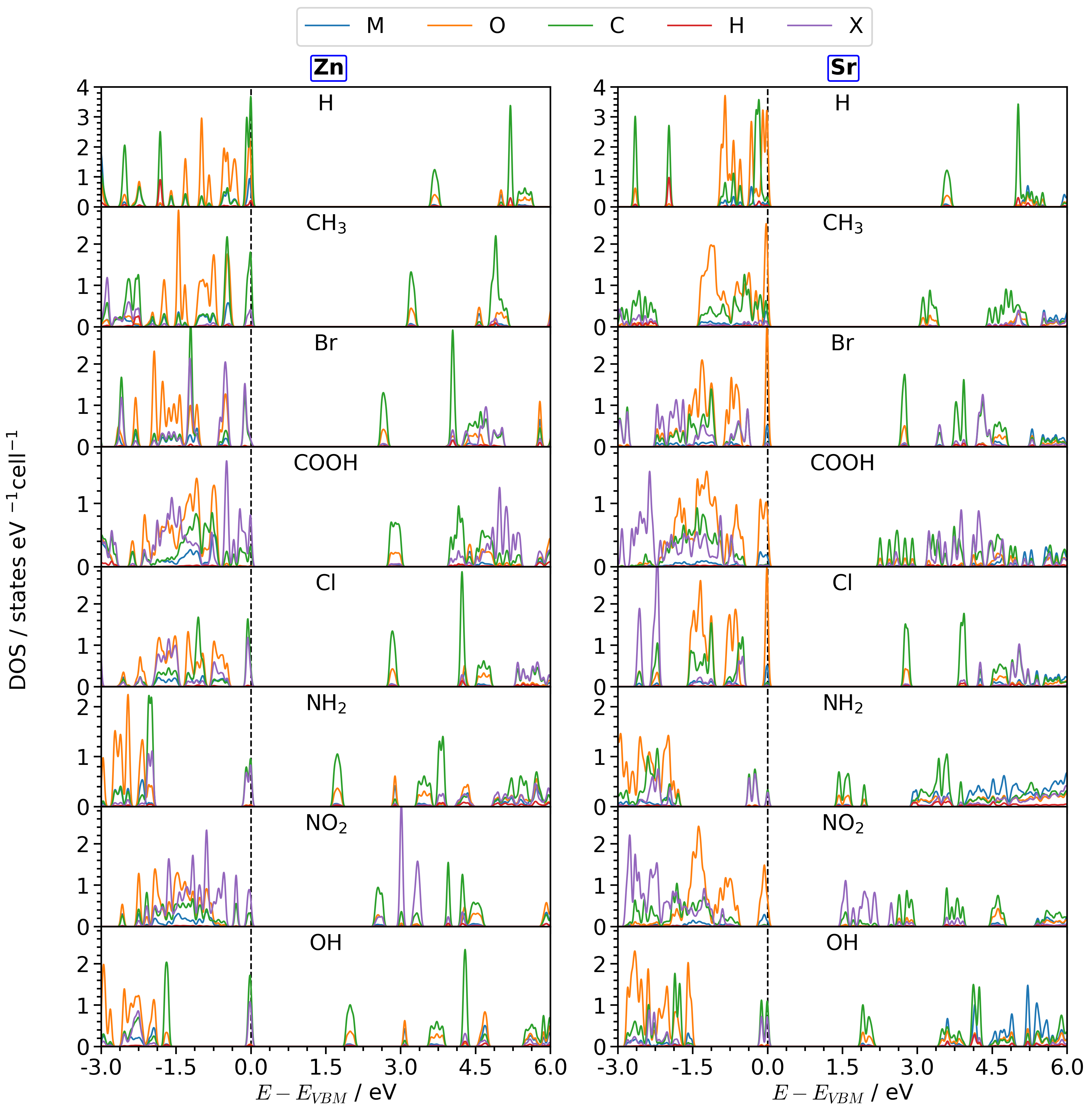}
    \caption{Calculated projected density of states for (left) Zn- and (right) Sr-terminated MOF-5 structures with all functionalization. The valance band maximum is set to $0$~eV.
    For better visualization, a one-dimensional Gaussian filter with $\sigma=0.05$~eV is applied.
    }
    \label{fig:PDOS}
\end{figure*}

The choice of the metal node does not affect the band gaps as much as the linker functionalization except for the COOH- and \ce{NO2}-terminated systems where the results obtained with different metallic species oscillate within a range of almost 1~eV (see Fig.~\ref{fig:Band_gap}). 
This behavior can be explained through the analysis of the projected density of states (pDOS) provided below.
The largest band gaps are obtained for the H-passivated MOF-5 followed by those with \ce{CH3}-functionalized ligands.
The halogen terminations, Cl and Br, give rise to MOFs with similar gaps, slightly lower than those with H and \ce{CH3} functionalization. 
The lowest band gaps are found with \ce{NH2} and OH groups bound to the linkers, in agreement with the findings obtained for other MOFs~\cite{sass+24ic}.
The physicochemical mechanism responsible for this behavior is the same as in Ref.~\citenum{sass+24ic} as discussed below in the analysis of the pDOS, see Fig.~\ref{fig:PDOS}, where the results obtained for the Zn- and Sr-based scaffold with all the considered linker functionalization are shown.
Zn is representative here of the smaller elements like Mg, Be, Ca, and Cd, while Sr and Ba which are metal ions with a larger van der Waals radius, which lead to substantially different band gaps upon COOH and \ce{NO2} functionalization.
The pDOS calculated for all structures are shown in the SI (Figures S2 — S6).

%%%%%%%%%%%%%%%%%%%%%%%%%%%%%%%%%%%%%%%%%%%%%%%%%%%%%%%%%%%%%%%%%%%%%
%% pDOS

We start this analysis from the Zn-based MOFs (Fig.~\ref{fig:PDOS}, left panel) and we notice at a glance that the conduction band minimum (CBM) is localized on the BDC linker regardless of its functionalization: the largest contribution to this state comes from the carbon backbone, but the influence of the oxygen atoms is non-negligible too.
On the other hand, the character of the valence band maximum (VBM) is much more affected by the ligand termination.
The pDOS of the H-passivated Zn-based MOF exhibits only one peak with H-character in the valence region, approximately 2~eV below the VBM.
The electronic states of the \ce{CH3} group slightly hybridize with the $\pi$ orbitals of the BDC linker even at the VBM, but they do not significantly affect the predominant carbon character of this state; additional contributions appear at lower energies in the valence band.
This hybridization with the BDC carbon $\pi$ states at the VBM becomes much more pronounced with Br, COOH, and Cl functionalization.
With \ce{NO2} this effect is even more pronounced: due to its remarkable electronegativity, the electronic states of this group strongly contribute to the gap region, both at the top of the valence band and with localized states right above the CBM.
It should be noted that the interaction with the BDC states does not occur primarily via electronic hybridization, as discussed above for the other terminations: rather, states localized on \ce{NO2} energetically fit in the pDOS of the MOF scaffold.
Finally, we notice that with \ce{NH2} and OH functionalization, the VBM is pinned in the mid-gap of the MOF by an orbital localized on the functional group. 
An analogous result was found in scandium terephthalates and its substituted variants~\cite{sass+24ic} suggesting that the influence of these two terminations on the electronic structure of MOFs with BDC linkers is not only independent of the metal node but even of the overall structure of the MOF.

This feature related to \ce{NH2} and OH functionalization is present also in the pDOS of MOF-5 with Sr (and Ba, see Fig.~S6) metal node (Fig.~\ref{fig:PDOS}, right panel) and can be related to the formation of hydrogen bonds discussed above.
However, this is the only clear similarity with the variants hosting smaller metallic species (Fig.~\ref{fig:PDOS}, left panel), at least in the valence region.
At a glance, we notice that in all Sr-based MOF-5 structures, except for those terminated with \ce{NH2} and OH, the VBM is dominated by oxygen contributions. 
In the H-passivated structure, the $\pi$ orbitals of the BDC linker are energetically close by and contribute equally to the top of the valence band.
A similar scenario is found also with \ce{CH3} termination, although the O-state at the VBM is more localized in this case.
The localized character of the oxygen states at the VBM is even more pronounced with the halogen terminations, Br and Cl, as well as with COOH and \ce{NO2}.
In the latter case, the contributions of the functional group pin the CBM but appear much deeper in the valence band compared to the counterparts with Zn as a metal node.
In all the other Sr-based MOF-5, the CBM has a carbon character, and the overall shape of the pDOS in the conduction region is much more similar to that of the Zn-based systems than in the valence band. 
We note in passing that the effect of these functional groups in complex metal-organic frameworks is more intricate than in pure carbon nanostructures, where the extension of the C-conjugated $\pi$ network is sufficiently large to prevent not only the presence of localized states on the functional groups but also remarkable hybridization, leaving only a rigid shift of the frontier states as the effect of edge functionalization~\cite{cocc+11jpcc,cocc+11jpcl}.
On the other hand, in small organic molecules such as oligothiophenes, the methyl termination has a similar effect as here on MOF-5 and is sufficient to model interactions with longer functionalizing alkyl chains~\cite{krum+21pccp}.

With the information gained from this analysis, we can go back to the trend obtained for the band gaps (Fig.~\ref{fig:Band_gap}) and explain in particular the large metal-node-dependent variations obtained with \ce{NO2} and COOH terminations.
From the pDOS shown in Fig.~\ref{fig:PDOS}, it is clear that the physical origin is different from all the other cases.
The reduction of the band gap in COOH-functionalized MOF-5 with larger metallic species is related to the presence of O-localized states at the top of the valence band (see Fig.~\ref{fig:PDOS}, right panel), a few hundred of meV above the upper threshold of occupied carbon states.
Some differences are seen in the unoccupied electronic region of these systems (see Fig.~\ref{fig:PDOS}, left and right panel) but no major differences appear.
With \ce{NO2} terminations, we see two different behaviors for the gap in the presence of small- or large-radius metallic nodes (see Fig.~\ref{fig:Band_gap}).
Considering Zn and Sr as representatives of each category, we notice that in the former case (Fig.~\ref{fig:PDOS}, left panel) the VBM is dominated by \ce{NO2} states and the CBM by $\pi^*$ states of the BDC backbone.
On the other hand, with Sr (or Ba) metal nodes, both frontier states are localized: the VBM on the oxygen atoms of the linker and the CBM on the functional group. 
This characteristic further reduces the gap and explains the trend seen in Fig.~\ref{fig:Band_gap}.

To fully understand this behavior, we still need to examine the contribution of the metallic species to the pDOS.
In the presence of Zn, no significant contributions from orbitals localized on the metal node are visible in the energy range plotted in Fig.~\ref{fig:PDOS}, left panel.
In contrast, Sr $d$-states are identifiable in the higher portion of the conduction region shown in Fig.~\ref{fig:PDOS}, right panel.
This is consistent with physicochemical intuition in three ways: (i) the larger number of electrons in Sr and Ba, leads to orbital contributions closer to the gap region; (ii) it increases screening, thus promoting an overall reduction of the band gap; (iii) it enhances electrostatic interactions with oxygen atoms which are close to these large metallic species due to their modified coordination (see Fig.~\ref{fig:struct}).

\section{Summary and Conclusions}
\label{sec:conclu}

In summary, we reported a first-principle analysis of the structural and electronic properties of MOF-5 variants with exchanged metal nodes and functionalized molecular linkers. We found that large metallic species and functional groups substantially affect the structure by altering the orientation of the ligand and consequently affecting the coordination of the metal ions. These characteristics impact the formation energy and the electronic properties of the MOFs.
Functionalization with COOH and OH groups promotes hydrogen bonds which enhance the stability of the systems regardless of the metal nodes, although a clear preference for alkali-earth over transition-metal species can be identified.
Partial charges are affected by functionalization, too: a large electronegativity in the terminating groups amplifies the ionic nature of the bonding to BDC.
A large tunability for the band gaps is found upon ligand functionalization. Single atoms or small groups like H, \ce{CH3}, Cl, and Br lead to the largest gaps, with essentially no variations depending on the choice of the metallic species. 
In contrast, with COOH and \ce{NO2}, a remarkable dependence on the metallic node is identified: larger species like Sr and Ba diminish the band-gap value due to the change in the metal-ion coordination induced by steric distortions. 
Finally, the electron-donating groups \ce{NH2} and OH give rise to the structures with the smallest band gaps due to the introduction of a localized orbital at the valence-band top.
This characteristic, which is expected to have a significant impact on the gas uptake and sensing properties of the corresponding MOFs, is promoted by the modified coordination of the large atomic species Sr and Ba compared to smaller elements.

In conclusion, this work shows the potential of high-throughput \textit{ab initio} simulations to unveil the structures and properties of functionalized MOFs efficiently. Our results demonstrate that rather than the chemical characteristics of specific metal species, the size of these atoms plays the most important role in the structural and electronic properties of the considered systems.
On the other hand, ligand functionalization with \ce{NH2} and OH groups appears to be ideally suited for MOFs targeting gas storage and sensing applications, due to the presence of a localized orbital at the valence-band top that they induce. 

\section*{Supplementary Material}
We report raw data of structural parameters, formation energies, partial charges, and band gaps for all considered compounds. Band structures and projected density of states of all calculated materials are reported, too.
	
\section*{Data Availability}
The data produced in this study are openly available in Zenodo at 10.5281/zenodo.10624825.

%% classes.  This should be given within the "acknowledgment"
\begin{acknowledgments}
This work was financed by the German Federal Ministry of Education and Research (Professorinnenprogramm III) and by the State of Lower Saxony (Professorinnen f\"ur Niedersachsen).
J.E. appreciates additional support from the Nagelschneider Stiftung and J.S.A. acknowledges funding from the Evonik Stiftung.
Computational resources were provided by the North-German Supercomputing Alliance (HLRN), project nic00069 and nic00084, and by the local high-performance computing cluster CARL at the University of Oldenburg, financed by the German Research Foundation (Project No. INST 184/157-1 FUGG) and by the Ministry of Science and Culture of the State of Lower Saxony.

\end{acknowledgments}

%\bibliography{references}

\begin{thebibliography}{58}%
\makeatletter
\providecommand \@ifxundefined [1]{%
 \@ifx{#1\undefined}
}%
\providecommand \@ifnum [1]{%
 \ifnum #1\expandafter \@firstoftwo
 \else \expandafter \@secondoftwo
 \fi
}%
\providecommand \@ifx [1]{%
 \ifx #1\expandafter \@firstoftwo
 \else \expandafter \@secondoftwo
 \fi
}%
\providecommand \natexlab [1]{#1}%
\providecommand \enquote  [1]{``#1''}%
\providecommand \bibnamefont  [1]{#1}%
\providecommand \bibfnamefont [1]{#1}%
\providecommand \citenamefont [1]{#1}%
\providecommand \href@noop [0]{\@secondoftwo}%
\providecommand \href [0]{\begingroup \@sanitize@url \@href}%
\providecommand \@href[1]{\@@startlink{#1}\@@href}%
\providecommand \@@href[1]{\endgroup#1\@@endlink}%
\providecommand \@sanitize@url [0]{\catcode `\\12\catcode `\$12\catcode
  `\&12\catcode `\#12\catcode `\^12\catcode `\_12\catcode `\%12\relax}%
\providecommand \@@startlink[1]{}%
\providecommand \@@endlink[0]{}%
\providecommand \url  [0]{\begingroup\@sanitize@url \@url }%
\providecommand \@url [1]{\endgroup\@href {#1}{\urlprefix }}%
\providecommand \urlprefix  [0]{URL }%
\providecommand \Eprint [0]{\href }%
\providecommand \doibase [0]{http://dx.doi.org/}%
\providecommand \selectlanguage [0]{\@gobble}%
\providecommand \bibinfo  [0]{\@secondoftwo}%
\providecommand \bibfield  [0]{\@secondoftwo}%
\providecommand \translation [1]{[#1]}%
\providecommand \BibitemOpen [0]{}%
\providecommand \bibitemStop [0]{}%
\providecommand \bibitemNoStop [0]{.\EOS\space}%
\providecommand \EOS [0]{\spacefactor3000\relax}%
\providecommand \BibitemShut  [1]{\csname bibitem#1\endcsname}%
\let\auto@bib@innerbib\@empty
%</preamble>
\bibitem [{\citenamefont {Furukawa}\ \emph {et~al.}(2013)\citenamefont
  {Furukawa}, \citenamefont {Cordova}, \citenamefont {O’Keeffe},\ and\
  \citenamefont {Yaghi}}]{furu+13sci}%
  \BibitemOpen
  \bibfield  {author} {\bibinfo {author} {\bibfnamefont {H.}~\bibnamefont
  {Furukawa}}, \bibinfo {author} {\bibfnamefont {K.~E.}\ \bibnamefont
  {Cordova}}, \bibinfo {author} {\bibfnamefont {M.}~\bibnamefont {O’Keeffe}},
  \ and\ \bibinfo {author} {\bibfnamefont {O.~M.}\ \bibnamefont {Yaghi}},\
  }\href@noop {} {\bibfield  {journal} {\bibinfo  {journal} {Science}\ }\textbf
  {\bibinfo {volume} {341}},\ \bibinfo {pages} {1230444} (\bibinfo {year}
  {2013})}\BibitemShut {NoStop}%
\bibitem [{\citenamefont {James}(2003)}]{jame+03csr}%
  \BibitemOpen
  \bibfield  {author} {\bibinfo {author} {\bibfnamefont {S.~L.}\ \bibnamefont
  {James}},\ }\href@noop {} {\bibfield  {journal} {\bibinfo  {journal}
  {Chem.~Soc.~Rev.}\ }\textbf {\bibinfo {volume} {32}},\ \bibinfo {pages} {276}
  (\bibinfo {year} {2003})}\BibitemShut {NoStop}%
\bibitem [{\citenamefont {Rosi}\ \emph {et~al.}(2003)\citenamefont {Rosi},
  \citenamefont {Eckert}, \citenamefont {Eddaoudi}, \citenamefont {Vodak},
  \citenamefont {Kim}, \citenamefont {O'Keeffe},\ and\ \citenamefont
  {Yaghi}}]{rosi+03sci}%
  \BibitemOpen
  \bibfield  {author} {\bibinfo {author} {\bibfnamefont {N.~L.}\ \bibnamefont
  {Rosi}}, \bibinfo {author} {\bibfnamefont {J.}~\bibnamefont {Eckert}},
  \bibinfo {author} {\bibfnamefont {M.}~\bibnamefont {Eddaoudi}}, \bibinfo
  {author} {\bibfnamefont {D.~T.}\ \bibnamefont {Vodak}}, \bibinfo {author}
  {\bibfnamefont {J.}~\bibnamefont {Kim}}, \bibinfo {author} {\bibfnamefont
  {M.}~\bibnamefont {O'Keeffe}}, \ and\ \bibinfo {author} {\bibfnamefont
  {O.~M.}\ \bibnamefont {Yaghi}},\ }\href {\doibase 10.1126/science.1083440}
  {\bibfield  {journal} {\bibinfo  {journal} {Science}\ }\textbf {\bibinfo
  {volume} {300}},\ \bibinfo {pages} {1127} (\bibinfo {year} {2003})},\ \Eprint
  {http://arxiv.org/abs/https://www.science.org/doi/pdf/10.1126/science.1083440}
  {https://www.science.org/doi/pdf/10.1126/science.1083440} \BibitemShut
  {NoStop}%
\bibitem [{\citenamefont {Koo}\ \emph {et~al.}(2019)\citenamefont {Koo},
  \citenamefont {Jang},\ and\ \citenamefont {Kim}}]{koo+19chem}%
  \BibitemOpen
  \bibfield  {author} {\bibinfo {author} {\bibfnamefont {W.-T.}\ \bibnamefont
  {Koo}}, \bibinfo {author} {\bibfnamefont {J.-S.}\ \bibnamefont {Jang}}, \
  and\ \bibinfo {author} {\bibfnamefont {I.-D.}\ \bibnamefont {Kim}},\ }\href
  {\doibase https://doi.org/10.1016/j.chempr.2019.04.013} {\bibfield  {journal}
  {\bibinfo  {journal} {Chem}\ }\textbf {\bibinfo {volume} {5}},\ \bibinfo
  {pages} {1938} (\bibinfo {year} {2019})}\BibitemShut {NoStop}%
\bibitem [{\citenamefont {Nguyen}\ \emph {et~al.}(2020)\citenamefont {Nguyen},
  \citenamefont {Lee}, \citenamefont {Doan}, \citenamefont {Nguyen},
  \citenamefont {Park}, \citenamefont {Kim},\ and\ \citenamefont
  {Phan}}]{nguy+20ass}%
  \BibitemOpen
  \bibfield  {author} {\bibinfo {author} {\bibfnamefont {D.-K.}\ \bibnamefont
  {Nguyen}}, \bibinfo {author} {\bibfnamefont {J.-H.}\ \bibnamefont {Lee}},
  \bibinfo {author} {\bibfnamefont {T.~L.-H.}\ \bibnamefont {Doan}}, \bibinfo
  {author} {\bibfnamefont {T.-B.}\ \bibnamefont {Nguyen}}, \bibinfo {author}
  {\bibfnamefont {S.}~\bibnamefont {Park}}, \bibinfo {author} {\bibfnamefont
  {S.~S.}\ \bibnamefont {Kim}}, \ and\ \bibinfo {author} {\bibfnamefont
  {B.~T.}\ \bibnamefont {Phan}},\ }\href {\doibase
  https://doi.org/10.1016/j.apsusc.2020.146487} {\bibfield  {journal} {\bibinfo
   {journal} {Appl.~Surf.~Sci.~}\ }\textbf {\bibinfo {volume} {523}},\ \bibinfo
  {pages} {146487} (\bibinfo {year} {2020})}\BibitemShut {NoStop}%
\bibitem [{\citenamefont {Ma}\ \emph {et~al.}(2021)\citenamefont {Ma},
  \citenamefont {Liu}, \citenamefont {Yang}, \citenamefont {Mao}, \citenamefont
  {Zheng},\ and\ \citenamefont {Jiang}}]{ma+21jacs}%
  \BibitemOpen
  \bibfield  {author} {\bibinfo {author} {\bibfnamefont {X.}~\bibnamefont
  {Ma}}, \bibinfo {author} {\bibfnamefont {H.}~\bibnamefont {Liu}}, \bibinfo
  {author} {\bibfnamefont {W.}~\bibnamefont {Yang}}, \bibinfo {author}
  {\bibfnamefont {G.}~\bibnamefont {Mao}}, \bibinfo {author} {\bibfnamefont
  {L.}~\bibnamefont {Zheng}}, \ and\ \bibinfo {author} {\bibfnamefont {H.-L.}\
  \bibnamefont {Jiang}},\ }\href {\doibase 10.1021/jacs.1c05032} {\bibfield
  {journal} {\bibinfo  {journal} {J.~Am.~Chem.~Soc.~}\ }\textbf {\bibinfo
  {volume} {143}},\ \bibinfo {pages} {12220} (\bibinfo {year}
  {2021})}\BibitemShut {NoStop}%
\bibitem [{\citenamefont {Wu}\ \emph {et~al.}(2020)\citenamefont {Wu},
  \citenamefont {Liu}, \citenamefont {Liu}, \citenamefont {Cheng},
  \citenamefont {Liu}, \citenamefont {Zeng}, \citenamefont {Shao},
  \citenamefont {Liang}, \citenamefont {Zhang}, \citenamefont {He},\ and\
  \citenamefont {Zhang}}]{wu+20ccr}%
  \BibitemOpen
  \bibfield  {author} {\bibinfo {author} {\bibfnamefont {T.}~\bibnamefont
  {Wu}}, \bibinfo {author} {\bibfnamefont {X.}~\bibnamefont {Liu}}, \bibinfo
  {author} {\bibfnamefont {Y.}~\bibnamefont {Liu}}, \bibinfo {author}
  {\bibfnamefont {M.}~\bibnamefont {Cheng}}, \bibinfo {author} {\bibfnamefont
  {Z.}~\bibnamefont {Liu}}, \bibinfo {author} {\bibfnamefont {G.}~\bibnamefont
  {Zeng}}, \bibinfo {author} {\bibfnamefont {B.}~\bibnamefont {Shao}}, \bibinfo
  {author} {\bibfnamefont {Q.}~\bibnamefont {Liang}}, \bibinfo {author}
  {\bibfnamefont {W.}~\bibnamefont {Zhang}}, \bibinfo {author} {\bibfnamefont
  {Q.}~\bibnamefont {He}}, \ and\ \bibinfo {author} {\bibfnamefont
  {W.}~\bibnamefont {Zhang}},\ }\href {\doibase
  https://doi.org/10.1016/j.ccr.2019.213097} {\bibfield  {journal} {\bibinfo
  {journal} {Coord.~Chem.~Rev.~}\ }\textbf {\bibinfo {volume} {403}},\ \bibinfo
  {pages} {213097} (\bibinfo {year} {2020})}\BibitemShut {NoStop}%
\bibitem [{\citenamefont {Pratik}\ \emph {et~al.}(2020)\citenamefont {Pratik},
  \citenamefont {Gagliardi},\ and\ \citenamefont {Cramer}}]{prat+20jpcc}%
  \BibitemOpen
  \bibfield  {author} {\bibinfo {author} {\bibfnamefont {S.~M.}\ \bibnamefont
  {Pratik}}, \bibinfo {author} {\bibfnamefont {L.}~\bibnamefont {Gagliardi}}, \
  and\ \bibinfo {author} {\bibfnamefont {C.~J.}\ \bibnamefont {Cramer}},\
  }\href {\doibase 10.1021/acs.jpcc.9b10834} {\bibfield  {journal} {\bibinfo
  {journal} {J.~Phys.~Chem.~C}\ }\textbf {\bibinfo {volume} {124}},\ \bibinfo
  {pages} {1878} (\bibinfo {year} {2020})}\BibitemShut {NoStop}%
\bibitem [{\citenamefont {Liu}\ \emph {et~al.}(2023)\citenamefont {Liu},
  \citenamefont {Wen}, \citenamefont {Xiao}, \citenamefont {Tan}, \citenamefont
  {Qin}, \citenamefont {Li}, \citenamefont {Bai}, \citenamefont {Xi},
  \citenamefont {Yang}, \citenamefont {Fang}, \citenamefont {Hu}, \citenamefont
  {Gu},\ and\ \citenamefont {Zhu}}]{liu+23nl}%
  \BibitemOpen
  \bibfield  {author} {\bibinfo {author} {\bibfnamefont {M.}~\bibnamefont
  {Liu}}, \bibinfo {author} {\bibfnamefont {J.}~\bibnamefont {Wen}}, \bibinfo
  {author} {\bibfnamefont {R.}~\bibnamefont {Xiao}}, \bibinfo {author}
  {\bibfnamefont {R.}~\bibnamefont {Tan}}, \bibinfo {author} {\bibfnamefont
  {Y.}~\bibnamefont {Qin}}, \bibinfo {author} {\bibfnamefont {J.}~\bibnamefont
  {Li}}, \bibinfo {author} {\bibfnamefont {Y.}~\bibnamefont {Bai}}, \bibinfo
  {author} {\bibfnamefont {M.}~\bibnamefont {Xi}}, \bibinfo {author}
  {\bibfnamefont {W.}~\bibnamefont {Yang}}, \bibinfo {author} {\bibfnamefont
  {Q.}~\bibnamefont {Fang}}, \bibinfo {author} {\bibfnamefont {L.}~\bibnamefont
  {Hu}}, \bibinfo {author} {\bibfnamefont {W.}~\bibnamefont {Gu}}, \ and\
  \bibinfo {author} {\bibfnamefont {C.}~\bibnamefont {Zhu}},\ }\href {\doibase
  10.1021/acs.nanolett.3c01650} {\bibfield  {journal} {\bibinfo  {journal}
  {Nano~Lett.~}\ }\textbf {\bibinfo {volume} {23}},\ \bibinfo {pages} {5358}
  (\bibinfo {year} {2023})}\BibitemShut {NoStop}%
\bibitem [{\citenamefont {Liu}\ \emph {et~al.}(2010)\citenamefont {Liu},
  \citenamefont {Shioyama}, \citenamefont {Jiang}, \citenamefont {Zhang},\ and\
  \citenamefont {Xu}}]{liu+10car}%
  \BibitemOpen
  \bibfield  {author} {\bibinfo {author} {\bibfnamefont {B.}~\bibnamefont
  {Liu}}, \bibinfo {author} {\bibfnamefont {H.}~\bibnamefont {Shioyama}},
  \bibinfo {author} {\bibfnamefont {H.}~\bibnamefont {Jiang}}, \bibinfo
  {author} {\bibfnamefont {X.}~\bibnamefont {Zhang}}, \ and\ \bibinfo {author}
  {\bibfnamefont {Q.}~\bibnamefont {Xu}},\ }\href {\doibase
  https://doi.org/10.1016/j.carbon.2009.09.061} {\bibfield  {journal} {\bibinfo
   {journal} {Carbon}\ }\textbf {\bibinfo {volume} {48}},\ \bibinfo {pages}
  {456} (\bibinfo {year} {2010})}\BibitemShut {NoStop}%
\bibitem [{\citenamefont {Wang}\ \emph {et~al.}(2016)\citenamefont {Wang},
  \citenamefont {Han}, \citenamefont {Feng}, \citenamefont {Zhou},
  \citenamefont {Qi},\ and\ \citenamefont {Wang}}]{wang+16ccr}%
  \BibitemOpen
  \bibfield  {author} {\bibinfo {author} {\bibfnamefont {L.}~\bibnamefont
  {Wang}}, \bibinfo {author} {\bibfnamefont {Y.}~\bibnamefont {Han}}, \bibinfo
  {author} {\bibfnamefont {X.}~\bibnamefont {Feng}}, \bibinfo {author}
  {\bibfnamefont {J.}~\bibnamefont {Zhou}}, \bibinfo {author} {\bibfnamefont
  {P.}~\bibnamefont {Qi}}, \ and\ \bibinfo {author} {\bibfnamefont
  {B.}~\bibnamefont {Wang}},\ }\href {\doibase
  https://doi.org/10.1016/j.ccr.2015.09.002} {\bibfield  {journal} {\bibinfo
  {journal} {Coord.~Chem.~Rev.~}\ }\textbf {\bibinfo {volume} {307}},\ \bibinfo
  {pages} {361} (\bibinfo {year} {2016})},\ \bibinfo {note} {chemistry and
  Applications of Metal Organic Frameworks}\BibitemShut {NoStop}%
\bibitem [{\citenamefont {Yuan}\ \emph {et~al.}(2020)\citenamefont {Yuan},
  \citenamefont {Huang}, \citenamefont {Huang}, \citenamefont {Sun},
  \citenamefont {Qin}, \citenamefont {Feng}, \citenamefont {Li}, \citenamefont
  {Zou}, \citenamefont {Cagin},\ and\ \citenamefont {Zhou}}]{yuan+10jacs}%
  \BibitemOpen
  \bibfield  {author} {\bibinfo {author} {\bibfnamefont {S.}~\bibnamefont
  {Yuan}}, \bibinfo {author} {\bibfnamefont {L.}~\bibnamefont {Huang}},
  \bibinfo {author} {\bibfnamefont {Z.}~\bibnamefont {Huang}}, \bibinfo
  {author} {\bibfnamefont {D.}~\bibnamefont {Sun}}, \bibinfo {author}
  {\bibfnamefont {J.-S.}\ \bibnamefont {Qin}}, \bibinfo {author} {\bibfnamefont
  {L.}~\bibnamefont {Feng}}, \bibinfo {author} {\bibfnamefont {J.}~\bibnamefont
  {Li}}, \bibinfo {author} {\bibfnamefont {X.}~\bibnamefont {Zou}}, \bibinfo
  {author} {\bibfnamefont {T.}~\bibnamefont {Cagin}}, \ and\ \bibinfo {author}
  {\bibfnamefont {H.-C.}\ \bibnamefont {Zhou}},\ }\href {\doibase
  10.1021/jacs.9b13072} {\bibfield  {journal} {\bibinfo  {journal}
  {J.~Am.~Chem.~Soc.~}\ }\textbf {\bibinfo {volume} {142}},\ \bibinfo {pages}
  {4732} (\bibinfo {year} {2020})}\BibitemShut {NoStop}%
\bibitem [{\citenamefont {Chen}\ \emph {et~al.}(2016)\citenamefont {Chen},
  \citenamefont {Madden}, \citenamefont {Pham}, \citenamefont {Forrest},
  \citenamefont {Kumar}, \citenamefont {Yang}, \citenamefont {Xue},
  \citenamefont {Space}, \citenamefont {Perry~IV}, \citenamefont {Zhang},
  \citenamefont {Chen},\ and\ \citenamefont {Zaworotko}}]{chen+16acie}%
  \BibitemOpen
  \bibfield  {author} {\bibinfo {author} {\bibfnamefont {K.-J.}\ \bibnamefont
  {Chen}}, \bibinfo {author} {\bibfnamefont {D.~G.}\ \bibnamefont {Madden}},
  \bibinfo {author} {\bibfnamefont {T.}~\bibnamefont {Pham}}, \bibinfo {author}
  {\bibfnamefont {K.~A.}\ \bibnamefont {Forrest}}, \bibinfo {author}
  {\bibfnamefont {A.}~\bibnamefont {Kumar}}, \bibinfo {author} {\bibfnamefont
  {Q.-Y.}\ \bibnamefont {Yang}}, \bibinfo {author} {\bibfnamefont
  {W.}~\bibnamefont {Xue}}, \bibinfo {author} {\bibfnamefont {B.}~\bibnamefont
  {Space}}, \bibinfo {author} {\bibfnamefont {J.~J.}\ \bibnamefont {Perry~IV}},
  \bibinfo {author} {\bibfnamefont {J.-P.}\ \bibnamefont {Zhang}}, \bibinfo
  {author} {\bibfnamefont {X.-M.}\ \bibnamefont {Chen}}, \ and\ \bibinfo
  {author} {\bibfnamefont {M.~J.}\ \bibnamefont {Zaworotko}},\ }\href {\doibase
  https://doi.org/10.1002/anie.201603934} {\bibfield  {journal} {\bibinfo
  {journal} {Angew.~Chem.~Int.~Ed.~}\ }\textbf {\bibinfo {volume} {55}},\
  \bibinfo {pages} {10268} (\bibinfo {year} {2016})}\BibitemShut {NoStop}%
\bibitem [{\citenamefont {Liang}\ \emph {et~al.}(2014)\citenamefont {Liang},
  \citenamefont {Chevreau}, \citenamefont {Ragon}, \citenamefont {Southon},
  \citenamefont {Peterson},\ and\ \citenamefont {D{'}Alessandro}}]{lian+14cec}%
  \BibitemOpen
  \bibfield  {author} {\bibinfo {author} {\bibfnamefont {W.}~\bibnamefont
  {Liang}}, \bibinfo {author} {\bibfnamefont {H.}~\bibnamefont {Chevreau}},
  \bibinfo {author} {\bibfnamefont {F.}~\bibnamefont {Ragon}}, \bibinfo
  {author} {\bibfnamefont {P.~D.}\ \bibnamefont {Southon}}, \bibinfo {author}
  {\bibfnamefont {V.~K.}\ \bibnamefont {Peterson}}, \ and\ \bibinfo {author}
  {\bibfnamefont {D.~M.}\ \bibnamefont {D{'}Alessandro}},\ }\href {\doibase
  10.1039/C4CE01031K} {\bibfield  {journal} {\bibinfo  {journal}
  {CrystEngComm}\ }\textbf {\bibinfo {volume} {16}},\ \bibinfo {pages} {6530}
  (\bibinfo {year} {2014})}\BibitemShut {NoStop}%
\bibitem [{\citenamefont {Du}\ \emph {et~al.}(2013)\citenamefont {Du},
  \citenamefont {Lu}, \citenamefont {Zheng}, \citenamefont {Wang},
  \citenamefont {Zheng}, \citenamefont {Pan}, \citenamefont {You},\ and\
  \citenamefont {Bai}}]{du+13jacs}%
  \BibitemOpen
  \bibfield  {author} {\bibinfo {author} {\bibfnamefont {L.}~\bibnamefont
  {Du}}, \bibinfo {author} {\bibfnamefont {Z.}~\bibnamefont {Lu}}, \bibinfo
  {author} {\bibfnamefont {K.}~\bibnamefont {Zheng}}, \bibinfo {author}
  {\bibfnamefont {J.}~\bibnamefont {Wang}}, \bibinfo {author} {\bibfnamefont
  {X.}~\bibnamefont {Zheng}}, \bibinfo {author} {\bibfnamefont
  {Y.}~\bibnamefont {Pan}}, \bibinfo {author} {\bibfnamefont {X.}~\bibnamefont
  {You}}, \ and\ \bibinfo {author} {\bibfnamefont {J.}~\bibnamefont {Bai}},\
  }\href {\doibase 10.1021/ja309992a} {\bibfield  {journal} {\bibinfo
  {journal} {J.~Am.~Chem.~Soc.~}\ }\textbf {\bibinfo {volume} {135}},\ \bibinfo
  {pages} {562} (\bibinfo {year} {2013})}\BibitemShut {NoStop}%
\bibitem [{\citenamefont {Assen}\ \emph {et~al.}(2015)\citenamefont {Assen},
  \citenamefont {Belmabkhout}, \citenamefont {Adil}, \citenamefont {Bhatt},
  \citenamefont {Xue}, \citenamefont {Jiang},\ and\ \citenamefont
  {Eddaoudi}}]{asse+15acie}%
  \BibitemOpen
  \bibfield  {author} {\bibinfo {author} {\bibfnamefont {A.~H.}\ \bibnamefont
  {Assen}}, \bibinfo {author} {\bibfnamefont {Y.}~\bibnamefont {Belmabkhout}},
  \bibinfo {author} {\bibfnamefont {K.}~\bibnamefont {Adil}}, \bibinfo {author}
  {\bibfnamefont {P.~M.}\ \bibnamefont {Bhatt}}, \bibinfo {author}
  {\bibfnamefont {D.-X.}\ \bibnamefont {Xue}}, \bibinfo {author} {\bibfnamefont
  {H.}~\bibnamefont {Jiang}}, \ and\ \bibinfo {author} {\bibfnamefont
  {M.}~\bibnamefont {Eddaoudi}},\ }\href {\doibase
  https://doi.org/10.1002/anie.201506345} {\bibfield  {journal} {\bibinfo
  {journal} {Angew.~Chem.~Int.~Ed.~}\ }\textbf {\bibinfo {volume} {54}},\
  \bibinfo {pages} {14353} (\bibinfo {year} {2015})}\BibitemShut {NoStop}%
\bibitem [{\citenamefont {Zhao}\ \emph {et~al.}(2011)\citenamefont {Zhao},
  \citenamefont {Timmons}, \citenamefont {Yuan},\ and\ \citenamefont
  {Zhou}}]{zhao+11acr}%
  \BibitemOpen
  \bibfield  {author} {\bibinfo {author} {\bibfnamefont {D.}~\bibnamefont
  {Zhao}}, \bibinfo {author} {\bibfnamefont {D.~J.}\ \bibnamefont {Timmons}},
  \bibinfo {author} {\bibfnamefont {D.}~\bibnamefont {Yuan}}, \ and\ \bibinfo
  {author} {\bibfnamefont {H.-C.}\ \bibnamefont {Zhou}},\ }\href {\doibase
  10.1021/ar100112y} {\bibfield  {journal} {\bibinfo  {journal}
  {Acc.~Chem.~Res.~}\ }\textbf {\bibinfo {volume} {44}},\ \bibinfo {pages}
  {123} (\bibinfo {year} {2011})}\BibitemShut {NoStop}%
\bibitem [{\citenamefont {Zhang}\ \emph {et~al.}(2017)\citenamefont {Zhang},
  \citenamefont {Zhou}, \citenamefont {Pham}, \citenamefont {Forrest},
  \citenamefont {Liu}, \citenamefont {He}, \citenamefont {Wu}, \citenamefont
  {Yildirim}, \citenamefont {Chen}, \citenamefont {Space}, \citenamefont {Pan},
  \citenamefont {Zaworotko},\ and\ \citenamefont {Bai}}]{zhan+17acie}%
  \BibitemOpen
  \bibfield  {author} {\bibinfo {author} {\bibfnamefont {M.}~\bibnamefont
  {Zhang}}, \bibinfo {author} {\bibfnamefont {W.}~\bibnamefont {Zhou}},
  \bibinfo {author} {\bibfnamefont {T.}~\bibnamefont {Pham}}, \bibinfo {author}
  {\bibfnamefont {K.~A.}\ \bibnamefont {Forrest}}, \bibinfo {author}
  {\bibfnamefont {W.}~\bibnamefont {Liu}}, \bibinfo {author} {\bibfnamefont
  {Y.}~\bibnamefont {He}}, \bibinfo {author} {\bibfnamefont {H.}~\bibnamefont
  {Wu}}, \bibinfo {author} {\bibfnamefont {T.}~\bibnamefont {Yildirim}},
  \bibinfo {author} {\bibfnamefont {B.}~\bibnamefont {Chen}}, \bibinfo {author}
  {\bibfnamefont {B.}~\bibnamefont {Space}}, \bibinfo {author} {\bibfnamefont
  {Y.}~\bibnamefont {Pan}}, \bibinfo {author} {\bibfnamefont {M.~J.}\
  \bibnamefont {Zaworotko}}, \ and\ \bibinfo {author} {\bibfnamefont
  {J.}~\bibnamefont {Bai}},\ }\href {\doibase
  https://doi.org/10.1002/anie.201704974} {\bibfield  {journal} {\bibinfo
  {journal} {Angew.~Chem.~Int.~Ed.~}\ }\textbf {\bibinfo {volume} {56}},\
  \bibinfo {pages} {11426} (\bibinfo {year} {2017})},\ \Eprint
  {http://arxiv.org/abs/https://onlinelibrary.wiley.com/doi/pdf/10.1002/anie.201704974}
  {https://onlinelibrary.wiley.com/doi/pdf/10.1002/anie.201704974} \BibitemShut
  {NoStop}%
\bibitem [{\citenamefont {Li}\ \emph {et~al.}(1999)\citenamefont {Li},
  \citenamefont {Eddaoudi}, \citenamefont {O'Keeffe},\ and\ \citenamefont
  {Yaghi}}]{Li+99nat}%
  \BibitemOpen
  \bibfield  {author} {\bibinfo {author} {\bibfnamefont {H.}~\bibnamefont
  {Li}}, \bibinfo {author} {\bibfnamefont {M.}~\bibnamefont {Eddaoudi}},
  \bibinfo {author} {\bibfnamefont {M.}~\bibnamefont {O'Keeffe}}, \ and\
  \bibinfo {author} {\bibfnamefont {O.~M.}\ \bibnamefont {Yaghi}},\ }\href
  {\doibase 10.1038/46248} {\bibfield  {journal} {\bibinfo  {journal} {Nature}\
  }\textbf {\bibinfo {volume} {402}},\ \bibinfo {pages} {276} (\bibinfo {year}
  {1999})}\BibitemShut {NoStop}%
\bibitem [{\citenamefont {Kaye}\ \emph {et~al.}(2007)\citenamefont {Kaye},
  \citenamefont {Dailly}, \citenamefont {Yaghi},\ and\ \citenamefont
  {Long}}]{kaye+07jacs}%
  \BibitemOpen
  \bibfield  {author} {\bibinfo {author} {\bibfnamefont {S.~S.}\ \bibnamefont
  {Kaye}}, \bibinfo {author} {\bibfnamefont {A.}~\bibnamefont {Dailly}},
  \bibinfo {author} {\bibfnamefont {O.~M.}\ \bibnamefont {Yaghi}}, \ and\
  \bibinfo {author} {\bibfnamefont {J.~R.}\ \bibnamefont {Long}},\ }\href
  {\doibase 10.1021/ja076877g} {\bibfield  {journal} {\bibinfo  {journal}
  {J.~Am.~Chem.~Soc.~}\ }\textbf {\bibinfo {volume} {129}},\ \bibinfo {pages}
  {14176} (\bibinfo {year} {2007})}\BibitemShut {NoStop}%
\bibitem [{\citenamefont {Li}\ \emph {et~al.}(2009)\citenamefont {Li},
  \citenamefont {Cheng}, \citenamefont {Zhao}, \citenamefont {Long},\ and\
  \citenamefont {Dong}}]{li+09ijhe}%
  \BibitemOpen
  \bibfield  {author} {\bibinfo {author} {\bibfnamefont {J.}~\bibnamefont
  {Li}}, \bibinfo {author} {\bibfnamefont {S.}~\bibnamefont {Cheng}}, \bibinfo
  {author} {\bibfnamefont {Q.}~\bibnamefont {Zhao}}, \bibinfo {author}
  {\bibfnamefont {P.}~\bibnamefont {Long}}, \ and\ \bibinfo {author}
  {\bibfnamefont {J.}~\bibnamefont {Dong}},\ }\href {\doibase
  https://doi.org/10.1016/j.ijhydene.2008.11.048} {\bibfield  {journal}
  {\bibinfo  {journal} {Int.~J.~Hydrog.~Energy}\ }\textbf {\bibinfo {volume}
  {34}},\ \bibinfo {pages} {1377} (\bibinfo {year} {2009})}\BibitemShut
  {NoStop}%
\bibitem [{\citenamefont {Yang}\ \emph {et~al.}(2012)\citenamefont {Yang},
  \citenamefont {Jung}, \citenamefont {Kim}, \citenamefont {Im},\ and\
  \citenamefont {Park}}]{yang+12ijhe}%
  \BibitemOpen
  \bibfield  {author} {\bibinfo {author} {\bibfnamefont {S.~J.}\ \bibnamefont
  {Yang}}, \bibinfo {author} {\bibfnamefont {H.}~\bibnamefont {Jung}}, \bibinfo
  {author} {\bibfnamefont {T.}~\bibnamefont {Kim}}, \bibinfo {author}
  {\bibfnamefont {J.~H.}\ \bibnamefont {Im}}, \ and\ \bibinfo {author}
  {\bibfnamefont {C.~R.}\ \bibnamefont {Park}},\ }\href {\doibase
  https://doi.org/10.1016/j.ijhydene.2011.12.163} {\bibfield  {journal}
  {\bibinfo  {journal} {Int.~J.~Hydrog.~Energy}\ }\textbf {\bibinfo {volume}
  {37}},\ \bibinfo {pages} {5777} (\bibinfo {year} {2012})}\BibitemShut
  {NoStop}%
\bibitem [{\citenamefont {Zhao}\ \emph {et~al.}(2013)\citenamefont {Zhao},
  \citenamefont {Ma}, \citenamefont {Kasik}, \citenamefont {Li},\ and\
  \citenamefont {Lin}}]{zhao+13iecr}%
  \BibitemOpen
  \bibfield  {author} {\bibinfo {author} {\bibfnamefont {Z.}~\bibnamefont
  {Zhao}}, \bibinfo {author} {\bibfnamefont {X.}~\bibnamefont {Ma}}, \bibinfo
  {author} {\bibfnamefont {A.}~\bibnamefont {Kasik}}, \bibinfo {author}
  {\bibfnamefont {Z.}~\bibnamefont {Li}}, \ and\ \bibinfo {author}
  {\bibfnamefont {Y.~S.}\ \bibnamefont {Lin}},\ }\href {\doibase
  10.1021/ie202777q} {\bibfield  {journal} {\bibinfo  {journal}
  {Ind.~Eng.~Chem.~Res.}\ }\textbf {\bibinfo {volume} {52}},\ \bibinfo {pages}
  {1102} (\bibinfo {year} {2013})}\BibitemShut {NoStop}%
\bibitem [{\citenamefont {Liu}\ \emph {et~al.}(2011)\citenamefont {Liu},
  \citenamefont {Xiang}, \citenamefont {Hu}, \citenamefont {Zheng},\ and\
  \citenamefont {Cao}}]{liu+11jmc}%
  \BibitemOpen
  \bibfield  {author} {\bibinfo {author} {\bibfnamefont {S.}~\bibnamefont
  {Liu}}, \bibinfo {author} {\bibfnamefont {Z.}~\bibnamefont {Xiang}}, \bibinfo
  {author} {\bibfnamefont {Z.}~\bibnamefont {Hu}}, \bibinfo {author}
  {\bibfnamefont {X.}~\bibnamefont {Zheng}}, \ and\ \bibinfo {author}
  {\bibfnamefont {D.}~\bibnamefont {Cao}},\ }\href {\doibase
  10.1039/C1JM10166H} {\bibfield  {journal} {\bibinfo  {journal}
  {J.~Mater.~Chem.}\ }\textbf {\bibinfo {volume} {21}},\ \bibinfo {pages}
  {6649} (\bibinfo {year} {2011})}\BibitemShut {NoStop}%
\bibitem [{\citenamefont {Katoch}\ \emph {et~al.}(2018)\citenamefont {Katoch},
  \citenamefont {Bhardwaj}, \citenamefont {Goyal},\ and\ \citenamefont
  {Gautam}}]{kato+18vac}%
  \BibitemOpen
  \bibfield  {author} {\bibinfo {author} {\bibfnamefont {A.}~\bibnamefont
  {Katoch}}, \bibinfo {author} {\bibfnamefont {R.}~\bibnamefont {Bhardwaj}},
  \bibinfo {author} {\bibfnamefont {N.}~\bibnamefont {Goyal}}, \ and\ \bibinfo
  {author} {\bibfnamefont {S.}~\bibnamefont {Gautam}},\ }\href {\doibase
  https://doi.org/10.1016/j.vacuum.2018.09.019} {\bibfield  {journal} {\bibinfo
   {journal} {Vacuum}\ }\textbf {\bibinfo {volume} {158}},\ \bibinfo {pages}
  {249} (\bibinfo {year} {2018})}\BibitemShut {NoStop}%
\bibitem [{\citenamefont {Botas}\ \emph {et~al.}(2010)\citenamefont {Botas},
  \citenamefont {Calleja}, \citenamefont {Sánchez-Sánchez},\ and\
  \citenamefont {Orcajo}}]{bota+10lan}%
  \BibitemOpen
  \bibfield  {author} {\bibinfo {author} {\bibfnamefont {J.~A.}\ \bibnamefont
  {Botas}}, \bibinfo {author} {\bibfnamefont {G.}~\bibnamefont {Calleja}},
  \bibinfo {author} {\bibfnamefont {M.}~\bibnamefont {Sánchez-Sánchez}}, \
  and\ \bibinfo {author} {\bibfnamefont {M.~G.}\ \bibnamefont {Orcajo}},\
  }\href {\doibase 10.1021/la100423a} {\bibfield  {journal} {\bibinfo
  {journal} {Langmuir}\ }\textbf {\bibinfo {volume} {26}},\ \bibinfo {pages}
  {5300} (\bibinfo {year} {2010})}\BibitemShut {NoStop}%
\bibitem [{\citenamefont {Yang}\ \emph {et~al.}(2014)\citenamefont {Yang},
  \citenamefont {Liu},\ and\ \citenamefont {Sun}}]{yang+14jssc}%
  \BibitemOpen
  \bibfield  {author} {\bibinfo {author} {\bibfnamefont {J.-M.}\ \bibnamefont
  {Yang}}, \bibinfo {author} {\bibfnamefont {Q.}~\bibnamefont {Liu}}, \ and\
  \bibinfo {author} {\bibfnamefont {W.-Y.}\ \bibnamefont {Sun}},\ }\href
  {\doibase https://doi.org/10.1016/j.jssc.2014.06.004} {\bibfield  {journal}
  {\bibinfo  {journal} {J.~Solid~State~Chem.}\ }\textbf {\bibinfo {volume}
  {218}},\ \bibinfo {pages} {50} (\bibinfo {year} {2014})}\BibitemShut
  {NoStop}%
\bibitem [{\citenamefont {Gangu}\ \emph {et~al.}(2022)\citenamefont {Gangu},
  \citenamefont {Maddila},\ and\ \citenamefont {Jonnalagadda}}]{gang+22rscadv}%
  \BibitemOpen
  \bibfield  {author} {\bibinfo {author} {\bibfnamefont {K.~K.}\ \bibnamefont
  {Gangu}}, \bibinfo {author} {\bibfnamefont {S.}~\bibnamefont {Maddila}}, \
  and\ \bibinfo {author} {\bibfnamefont {S.~B.}\ \bibnamefont {Jonnalagadda}},\
  }\href {\doibase 10.1039/D2RA01505F} {\bibfield  {journal} {\bibinfo
  {journal} {RSC Adv.}\ }\textbf {\bibinfo {volume} {12}},\ \bibinfo {pages}
  {14282} (\bibinfo {year} {2022})}\BibitemShut {NoStop}%
\bibitem [{\citenamefont {Canepa}\ \emph {et~al.}(2013)\citenamefont {Canepa},
  \citenamefont {Arter}, \citenamefont {Conwill}, \citenamefont {Johnson},
  \citenamefont {Shoemaker}, \citenamefont {Soliman},\ and\ \citenamefont
  {Thonhauser}}]{cane+13jmca}%
  \BibitemOpen
  \bibfield  {author} {\bibinfo {author} {\bibfnamefont {P.}~\bibnamefont
  {Canepa}}, \bibinfo {author} {\bibfnamefont {C.~A.}\ \bibnamefont {Arter}},
  \bibinfo {author} {\bibfnamefont {E.~M.}\ \bibnamefont {Conwill}}, \bibinfo
  {author} {\bibfnamefont {D.~H.}\ \bibnamefont {Johnson}}, \bibinfo {author}
  {\bibfnamefont {B.~A.}\ \bibnamefont {Shoemaker}}, \bibinfo {author}
  {\bibfnamefont {K.~Z.}\ \bibnamefont {Soliman}}, \ and\ \bibinfo {author}
  {\bibfnamefont {T.}~\bibnamefont {Thonhauser}},\ }\href {\doibase
  10.1039/C3TA12395B} {\bibfield  {journal} {\bibinfo  {journal}
  {J.~Mater.~Chem.~A}\ }\textbf {\bibinfo {volume} {1}},\ \bibinfo {pages}
  {13597} (\bibinfo {year} {2013})}\BibitemShut {NoStop}%
\bibitem [{\citenamefont {Rosen}\ \emph {et~al.}(2019)\citenamefont {Rosen},
  \citenamefont {Notestein},\ and\ \citenamefont {Snurr}}]{rose+19jcc}%
  \BibitemOpen
  \bibfield  {author} {\bibinfo {author} {\bibfnamefont {A.~S.}\ \bibnamefont
  {Rosen}}, \bibinfo {author} {\bibfnamefont {J.~M.}\ \bibnamefont
  {Notestein}}, \ and\ \bibinfo {author} {\bibfnamefont {R.~Q.}\ \bibnamefont
  {Snurr}},\ }\href {\doibase https://doi.org/10.1002/jcc.25787} {\bibfield
  {journal} {\bibinfo  {journal} {J.~Comput.~Chem.~}\ }\textbf {\bibinfo
  {volume} {40}},\ \bibinfo {pages} {1305} (\bibinfo {year}
  {2019})}\BibitemShut {NoStop}%
\bibitem [{\citenamefont {Saßnick}\ \emph {et~al.}(2024)\citenamefont
  {Saßnick}, \citenamefont {Machado Ferreira De~Araujo}, \citenamefont
  {Edzards},\ and\ \citenamefont {Cocchi}}]{sass+24ic}%
  \BibitemOpen
  \bibfield  {author} {\bibinfo {author} {\bibfnamefont {H.-D.}\ \bibnamefont
  {Saßnick}}, \bibinfo {author} {\bibfnamefont {F.}~\bibnamefont {Machado
  Ferreira De~Araujo}}, \bibinfo {author} {\bibfnamefont {J.}~\bibnamefont
  {Edzards}}, \ and\ \bibinfo {author} {\bibfnamefont {C.}~\bibnamefont
  {Cocchi}},\ }\href {\doibase 10.1021/acs.inorgchem.3c03945} {\bibfield
  {journal} {\bibinfo  {journal} {Inorg.~Chem.~}\ }\textbf {\bibinfo {volume}
  {63}},\ \bibinfo {pages} {2098} (\bibinfo {year} {2024})}\BibitemShut
  {NoStop}%
\bibitem [{\citenamefont {Saßnick}\ and\ \citenamefont
  {Cocchi}(2022)}]{Sassnick2022}%
  \BibitemOpen
  \bibfield  {author} {\bibinfo {author} {\bibfnamefont {H.-D.}\ \bibnamefont
  {Saßnick}}\ and\ \bibinfo {author} {\bibfnamefont {C.}~\bibnamefont
  {Cocchi}},\ }\href {\doibase 10.1063/5.0082710} {\bibfield  {journal}
  {\bibinfo  {journal} {J.~Chem.~Phys.~}\ }\textbf {\bibinfo {volume} {156}},\
  \bibinfo {pages} {104108} (\bibinfo {year} {2022})},\ \Eprint
  {http://arxiv.org/abs/https://doi.org/10.1063/5.0082710}
  {https://doi.org/10.1063/5.0082710} \BibitemShut {NoStop}%
\bibitem [{\citenamefont {Momma}\ and\ \citenamefont
  {Izumi}(2011)}]{Momma2011}%
  \BibitemOpen
  \bibfield  {author} {\bibinfo {author} {\bibfnamefont {K.}~\bibnamefont
  {Momma}}\ and\ \bibinfo {author} {\bibfnamefont {F.}~\bibnamefont {Izumi}},\
  }\href {\doibase 10.1107/S0021889811038970} {\bibfield  {journal} {\bibinfo
  {journal} {J.~Appl.~Cryst.~}\ }\textbf {\bibinfo {volume} {44}},\ \bibinfo
  {pages} {1272} (\bibinfo {year} {2011})}\BibitemShut {NoStop}%
\bibitem [{\citenamefont {Butler}\ \emph {et~al.}(2014)\citenamefont {Butler},
  \citenamefont {Hendon},\ and\ \citenamefont {Walsh}}]{Butler2014}%
  \BibitemOpen
  \bibfield  {author} {\bibinfo {author} {\bibfnamefont {K.~T.}\ \bibnamefont
  {Butler}}, \bibinfo {author} {\bibfnamefont {C.~H.}\ \bibnamefont {Hendon}},
  \ and\ \bibinfo {author} {\bibfnamefont {A.}~\bibnamefont {Walsh}},\ }\href
  {\doibase 10.1021/ja4110073} {\bibfield  {journal} {\bibinfo  {journal}
  {J.~Am.~Chem.~Soc.~}\ }\textbf {\bibinfo {volume} {136}},\ \bibinfo {pages}
  {2703} (\bibinfo {year} {2014})},\ \Eprint
  {http://arxiv.org/abs/https://doi.org/10.1021/ja4110073}
  {https://doi.org/10.1021/ja4110073} \BibitemShut {NoStop}%
\bibitem [{\citenamefont {Butler}\ \emph {et~al.}(2015)\citenamefont {Butler},
  \citenamefont {Hendon},\ and\ \citenamefont {Walsh}}]{Butler2020}%
  \BibitemOpen
  \bibfield  {author} {\bibinfo {author} {\bibfnamefont {K.~T.}\ \bibnamefont
  {Butler}}, \bibinfo {author} {\bibfnamefont {C.~H.}\ \bibnamefont {Hendon}},
  \ and\ \bibinfo {author} {\bibfnamefont {A.}~\bibnamefont {Walsh}},\ }\href
  {https://github.com/WMD-group/Crystal_structures/tree/master/MOFs} {\enquote
  {\bibinfo {title} {Crystal structures},}\ } (\bibinfo {year}
  {2015})\BibitemShut {NoStop}%
\bibitem [{\citenamefont {Cordero}\ \emph {et~al.}(2008)\citenamefont
  {Cordero}, \citenamefont {Gómez}, \citenamefont {Platero-Prats},
  \citenamefont {Revés}, \citenamefont {Echeverría}, \citenamefont
  {Cremades}, \citenamefont {Barragán},\ and\ \citenamefont
  {Alvarez}}]{Cordero2008}%
  \BibitemOpen
  \bibfield  {author} {\bibinfo {author} {\bibfnamefont {B.}~\bibnamefont
  {Cordero}}, \bibinfo {author} {\bibfnamefont {V.}~\bibnamefont {Gómez}},
  \bibinfo {author} {\bibfnamefont {A.~E.}\ \bibnamefont {Platero-Prats}},
  \bibinfo {author} {\bibfnamefont {M.}~\bibnamefont {Revés}}, \bibinfo
  {author} {\bibfnamefont {J.}~\bibnamefont {Echeverría}}, \bibinfo {author}
  {\bibfnamefont {E.}~\bibnamefont {Cremades}}, \bibinfo {author}
  {\bibfnamefont {F.}~\bibnamefont {Barragán}}, \ and\ \bibinfo {author}
  {\bibfnamefont {S.}~\bibnamefont {Alvarez}},\ }\href {\doibase
  10.1039/B801115J} {\bibfield  {journal} {\bibinfo  {journal} {Dalton Trans.}\
  ,\ \bibinfo {pages} {2832}} (\bibinfo {year} {2008})}\BibitemShut {NoStop}%
\bibitem [{\citenamefont {Haynes}(2014)}]{Haynes2014}%
  \BibitemOpen
  \bibfield  {author} {\bibinfo {author} {\bibfnamefont {W.}~\bibnamefont
  {Haynes}},\ }\href@noop {} {\emph {\bibinfo {title} {CRC Handbook of
  Chemistry and Physics}}},\ \bibinfo {edition} {95th}\ ed.\ (\bibinfo
  {publisher} {CRC Press},\ \bibinfo {year} {2014})\BibitemShut {NoStop}%
\bibitem [{\citenamefont {K\"uhne}\ \emph {et~al.}(2020)\citenamefont
  {K\"uhne}, \citenamefont {Iannuzzi}, \citenamefont {Del~Ben}, \citenamefont
  {Rybkin}, \citenamefont {Seewald}, \citenamefont {Stein}, \citenamefont
  {Laino}, \citenamefont {Khaliullin}, \citenamefont {Schütt}, \citenamefont
  {Schiffmann}, \citenamefont {Golze}, \citenamefont {Wilhelm}, \citenamefont
  {Chulkov}, \citenamefont {Bani-Hashemian}, \citenamefont {Weber},
  \citenamefont {Borštnik}, \citenamefont {Taillefumier}, \citenamefont
  {Jakobovits}, \citenamefont {Lazzaro}, \citenamefont {Pabst}, \citenamefont
  {Müller}, \citenamefont {Schade}, \citenamefont {Guidon}, \citenamefont
  {Andermatt}, \citenamefont {Holmberg}, \citenamefont {Schenter},
  \citenamefont {Hehn}, \citenamefont {Bussy}, \citenamefont {Belleflamme},
  \citenamefont {Tabacchi}, \citenamefont {Glöß}, \citenamefont {Lass},
  \citenamefont {Bethune}, \citenamefont {Mundy}, \citenamefont {Plessl},
  \citenamefont {Watkins}, \citenamefont {VandeVondele}, \citenamefont
  {Krack},\ and\ \citenamefont {Hutter}}]{Kuehne2020}%
  \BibitemOpen
  \bibfield  {author} {\bibinfo {author} {\bibfnamefont {T.~D.}\ \bibnamefont
  {K\"uhne}}, \bibinfo {author} {\bibfnamefont {M.}~\bibnamefont {Iannuzzi}},
  \bibinfo {author} {\bibfnamefont {M.}~\bibnamefont {Del~Ben}}, \bibinfo
  {author} {\bibfnamefont {V.~V.}\ \bibnamefont {Rybkin}}, \bibinfo {author}
  {\bibfnamefont {P.}~\bibnamefont {Seewald}}, \bibinfo {author} {\bibfnamefont
  {F.}~\bibnamefont {Stein}}, \bibinfo {author} {\bibfnamefont
  {T.}~\bibnamefont {Laino}}, \bibinfo {author} {\bibfnamefont {R.~Z.}\
  \bibnamefont {Khaliullin}}, \bibinfo {author} {\bibfnamefont
  {O.}~\bibnamefont {Schütt}}, \bibinfo {author} {\bibfnamefont
  {F.}~\bibnamefont {Schiffmann}}, \bibinfo {author} {\bibfnamefont
  {D.}~\bibnamefont {Golze}}, \bibinfo {author} {\bibfnamefont
  {J.}~\bibnamefont {Wilhelm}}, \bibinfo {author} {\bibfnamefont
  {S.}~\bibnamefont {Chulkov}}, \bibinfo {author} {\bibfnamefont {M.~H.}\
  \bibnamefont {Bani-Hashemian}}, \bibinfo {author} {\bibfnamefont
  {V.}~\bibnamefont {Weber}}, \bibinfo {author} {\bibfnamefont
  {U.}~\bibnamefont {Borštnik}}, \bibinfo {author} {\bibfnamefont
  {M.}~\bibnamefont {Taillefumier}}, \bibinfo {author} {\bibfnamefont {A.~S.}\
  \bibnamefont {Jakobovits}}, \bibinfo {author} {\bibfnamefont
  {A.}~\bibnamefont {Lazzaro}}, \bibinfo {author} {\bibfnamefont
  {H.}~\bibnamefont {Pabst}}, \bibinfo {author} {\bibfnamefont
  {T.}~\bibnamefont {Müller}}, \bibinfo {author} {\bibfnamefont
  {R.}~\bibnamefont {Schade}}, \bibinfo {author} {\bibfnamefont
  {M.}~\bibnamefont {Guidon}}, \bibinfo {author} {\bibfnamefont
  {S.}~\bibnamefont {Andermatt}}, \bibinfo {author} {\bibfnamefont
  {N.}~\bibnamefont {Holmberg}}, \bibinfo {author} {\bibfnamefont {G.~K.}\
  \bibnamefont {Schenter}}, \bibinfo {author} {\bibfnamefont {A.}~\bibnamefont
  {Hehn}}, \bibinfo {author} {\bibfnamefont {A.}~\bibnamefont {Bussy}},
  \bibinfo {author} {\bibfnamefont {F.}~\bibnamefont {Belleflamme}}, \bibinfo
  {author} {\bibfnamefont {G.}~\bibnamefont {Tabacchi}}, \bibinfo {author}
  {\bibfnamefont {A.}~\bibnamefont {Glöß}}, \bibinfo {author} {\bibfnamefont
  {M.}~\bibnamefont {Lass}}, \bibinfo {author} {\bibfnamefont {I.}~\bibnamefont
  {Bethune}}, \bibinfo {author} {\bibfnamefont {C.~J.}\ \bibnamefont {Mundy}},
  \bibinfo {author} {\bibfnamefont {C.}~\bibnamefont {Plessl}}, \bibinfo
  {author} {\bibfnamefont {M.}~\bibnamefont {Watkins}}, \bibinfo {author}
  {\bibfnamefont {J.}~\bibnamefont {VandeVondele}}, \bibinfo {author}
  {\bibfnamefont {M.}~\bibnamefont {Krack}}, \ and\ \bibinfo {author}
  {\bibfnamefont {J.}~\bibnamefont {Hutter}},\ }\href {\doibase
  10.1063/5.0007045} {\bibfield  {journal} {\bibinfo  {journal}
  {J.~Chem.~Phys.~}\ }\textbf {\bibinfo {volume} {152}},\ \bibinfo {pages}
  {194103} (\bibinfo {year} {2020})}\BibitemShut {NoStop}%
\bibitem [{\citenamefont {Goedecker}\ \emph {et~al.}(1996)\citenamefont
  {Goedecker}, \citenamefont {Teter},\ and\ \citenamefont
  {Hutter}}]{Goedecker1996}%
  \BibitemOpen
  \bibfield  {author} {\bibinfo {author} {\bibfnamefont {S.}~\bibnamefont
  {Goedecker}}, \bibinfo {author} {\bibfnamefont {M.}~\bibnamefont {Teter}}, \
  and\ \bibinfo {author} {\bibfnamefont {J.}~\bibnamefont {Hutter}},\ }\href
  {\doibase 10.1103/PhysRevB.54.1703} {\bibfield  {journal} {\bibinfo
  {journal} {Phys.~Rev.~B}\ }\textbf {\bibinfo {volume} {54}},\ \bibinfo
  {pages} {1703} (\bibinfo {year} {1996})}\BibitemShut {NoStop}%
\bibitem [{\citenamefont {Perdew}\ \emph {et~al.}(1996)\citenamefont {Perdew},
  \citenamefont {Burke},\ and\ \citenamefont {Ernzerhof}}]{perd+96prl}%
  \BibitemOpen
  \bibfield  {author} {\bibinfo {author} {\bibfnamefont {J.~P.}\ \bibnamefont
  {Perdew}}, \bibinfo {author} {\bibfnamefont {K.}~\bibnamefont {Burke}}, \
  and\ \bibinfo {author} {\bibfnamefont {M.}~\bibnamefont {Ernzerhof}},\
  }\href@noop {} {\bibfield  {journal} {\bibinfo  {journal}
  {Phys.~Rev.~Lett.~}\ }\textbf {\bibinfo {volume} {77}},\ \bibinfo {pages}
  {3865} (\bibinfo {year} {1996})}\BibitemShut {NoStop}%
\bibitem [{\citenamefont {Grimme}\ \emph {et~al.}(2010)\citenamefont {Grimme},
  \citenamefont {Antony}, \citenamefont {Ehrlich},\ and\ \citenamefont
  {Krieg}}]{Grimme2010}%
  \BibitemOpen
  \bibfield  {author} {\bibinfo {author} {\bibfnamefont {S.}~\bibnamefont
  {Grimme}}, \bibinfo {author} {\bibfnamefont {J.}~\bibnamefont {Antony}},
  \bibinfo {author} {\bibfnamefont {S.}~\bibnamefont {Ehrlich}}, \ and\
  \bibinfo {author} {\bibfnamefont {H.}~\bibnamefont {Krieg}},\ }\href
  {\doibase 10.1063/1.3382344} {\bibfield  {journal} {\bibinfo  {journal}
  {J.~Chem.~Phys.~}\ }\textbf {\bibinfo {volume} {132}},\ \bibinfo {pages}
  {154104} (\bibinfo {year} {2010})}\BibitemShut {NoStop}%
\bibitem [{\citenamefont {Jensen}(2007)}]{Jensen2007}%
  \BibitemOpen
  \bibfield  {author} {\bibinfo {author} {\bibfnamefont {F.}~\bibnamefont
  {Jensen}},\ }\href@noop {} {\emph {\bibinfo {title} {Introduction to
  Computational Chemistry}}},\ \bibinfo {edition} {2nd}\ ed.\ (\bibinfo
  {publisher} {Wiley},\ \bibinfo {year} {2007})\BibitemShut {NoStop}%
\bibitem [{\citenamefont {Monkhorst}\ and\ \citenamefont
  {Pack}(1976)}]{Monkhorst1976}%
  \BibitemOpen
  \bibfield  {author} {\bibinfo {author} {\bibfnamefont {H.~J.}\ \bibnamefont
  {Monkhorst}}\ and\ \bibinfo {author} {\bibfnamefont {J.~D.}\ \bibnamefont
  {Pack}},\ }\href {\doibase 10.1103/PhysRevB.13.5188} {\bibfield  {journal}
  {\bibinfo  {journal} {Phys.~Rev.~B}\ }\textbf {\bibinfo {volume} {13}},\
  \bibinfo {pages} {5188} (\bibinfo {year} {1976})}\BibitemShut {NoStop}%
\bibitem [{\citenamefont {Otero-de-la Roza}\ and\ \citenamefont
  {Luaña}(2014)}]{otero2014}%
  \BibitemOpen
  \bibfield  {author} {\bibinfo {author} {\bibfnamefont {A.~a. E. R.~J.}\
  \bibnamefont {Otero-de-la Roza}}\ and\ \bibinfo {author} {\bibfnamefont
  {V.}~\bibnamefont {Luaña}},\ }\href {\doibase
  https://doi.org/10.1016/j.cpc.2013.10.026} {\bibfield  {journal} {\bibinfo
  {journal} {Comput.~Phys.~Commun.~}\ }\textbf {\bibinfo {volume} {185}},\
  \bibinfo {pages} {1007} (\bibinfo {year} {2014})}\BibitemShut {NoStop}%
\bibitem [{\citenamefont {Bader}\ and\ \citenamefont {Zou}(1992)}]{Bader92cpl}%
  \BibitemOpen
  \bibfield  {author} {\bibinfo {author} {\bibfnamefont {R.}~\bibnamefont
  {Bader}}\ and\ \bibinfo {author} {\bibfnamefont {P.}~\bibnamefont {Zou}},\
  }\href {\doibase https://doi.org/10.1016/0009 -2614(92)85367-J} {\bibfield
  {journal} {\bibinfo  {journal} {Chem.~Phys.~Lett.~}\ }\textbf {\bibinfo
  {volume} {191}},\ \bibinfo {pages} {54} (\bibinfo {year} {1992})}\BibitemShut
  {NoStop}%
\bibitem [{\citenamefont {Yu}\ and\ \citenamefont {Trinkle}(2011)}]{Yu2011}%
  \BibitemOpen
  \bibfield  {author} {\bibinfo {author} {\bibfnamefont {M.}~\bibnamefont
  {Yu}}\ and\ \bibinfo {author} {\bibfnamefont {D.~R.}\ \bibnamefont
  {Trinkle}},\ }\href {\doibase 10.1063/1.3553716} {\bibfield  {journal}
  {\bibinfo  {journal} {J.~Chem.~Phys.~}\ }\textbf {\bibinfo {volume} {134}},\
  \bibinfo {pages} {064111} (\bibinfo {year} {2011})}\BibitemShut {NoStop}%
\bibitem [{\citenamefont {Shannon}(1976)}]{Shannon1976}%
  \BibitemOpen
  \bibfield  {author} {\bibinfo {author} {\bibfnamefont {R.~D.}\ \bibnamefont
  {Shannon}},\ }\href {\doibase 10.1107/S0567739476001551} {\bibfield
  {journal} {\bibinfo  {journal} {Acta Crystallogr.~A}\ }\textbf {\bibinfo
  {volume} {32}},\ \bibinfo {pages} {751} (\bibinfo {year} {1976})}\BibitemShut
  {NoStop}%
\bibitem [{\citenamefont {Yang}\ \emph {et~al.}(2011)\citenamefont {Yang},
  \citenamefont {Vajeeston}, \citenamefont {Ravindran}, \citenamefont
  {Fjellv\r{a}g},\ and\ \citenamefont {Tilset}}]{Yang2011}%
  \BibitemOpen
  \bibfield  {author} {\bibinfo {author} {\bibfnamefont {L.-M.}\ \bibnamefont
  {Yang}}, \bibinfo {author} {\bibfnamefont {P.}~\bibnamefont {Vajeeston}},
  \bibinfo {author} {\bibfnamefont {P.}~\bibnamefont {Ravindran}}, \bibinfo
  {author} {\bibfnamefont {H.}~\bibnamefont {Fjellv\r{a}g}}, \ and\ \bibinfo
  {author} {\bibfnamefont {M.}~\bibnamefont {Tilset}},\ }\href {\doibase
  10.1039/C0CP02944K} {\bibfield  {journal} {\bibinfo  {journal}
  {Phys.~Chem.~Chem.~Phys.~}\ }\textbf {\bibinfo {volume} {13}},\ \bibinfo
  {pages} {10191} (\bibinfo {year} {2011})}\BibitemShut {NoStop}%
\bibitem [{\citenamefont {Yang}\ \emph {et~al.}(2010)\citenamefont {Yang},
  \citenamefont {Vajeeston}, \citenamefont {Ravindran}, \citenamefont
  {Fjellv\r{a}g},\ and\ \citenamefont {Tilset}}]{Yang2010}%
  \BibitemOpen
  \bibfield  {author} {\bibinfo {author} {\bibfnamefont {L.-M.}\ \bibnamefont
  {Yang}}, \bibinfo {author} {\bibfnamefont {P.}~\bibnamefont {Vajeeston}},
  \bibinfo {author} {\bibfnamefont {P.}~\bibnamefont {Ravindran}}, \bibinfo
  {author} {\bibfnamefont {H.}~\bibnamefont {Fjellv\r{a}g}}, \ and\ \bibinfo
  {author} {\bibfnamefont {M.}~\bibnamefont {Tilset}},\ }\href {\doibase
  10.1021/ic100694w} {\bibfield  {journal} {\bibinfo  {journal}
  {Inorg.~Chem.~}\ }\textbf {\bibinfo {volume} {49}},\ \bibinfo {pages} {10283}
  (\bibinfo {year} {2010})}\BibitemShut {NoStop}%
\bibitem [{\citenamefont {Srepusharawoot}\ \emph {et~al.}(2008)\citenamefont
  {Srepusharawoot}, \citenamefont {Ara\'{u}jo}, \citenamefont {Blomqvist},
  \citenamefont {Scheicher},\ and\ \citenamefont {Ahuja}}]{Srepusharawoot2008}%
  \BibitemOpen
  \bibfield  {author} {\bibinfo {author} {\bibfnamefont {P.}~\bibnamefont
  {Srepusharawoot}}, \bibinfo {author} {\bibfnamefont {C.~M.}\ \bibnamefont
  {Ara\'{u}jo}}, \bibinfo {author} {\bibfnamefont {A.}~\bibnamefont
  {Blomqvist}}, \bibinfo {author} {\bibfnamefont {R.~H.}\ \bibnamefont
  {Scheicher}}, \ and\ \bibinfo {author} {\bibfnamefont {R.}~\bibnamefont
  {Ahuja}},\ }\href {\doibase 10.1063/1.2997377} {\bibfield  {journal}
  {\bibinfo  {journal} {J.~Chem.~Phys.~}\ }\textbf {\bibinfo {volume} {129}},\
  \bibinfo {pages} {164104} (\bibinfo {year} {2008})}\BibitemShut {NoStop}%
\bibitem [{\citenamefont {Rowsell}\ \emph {et~al.}(2004)\citenamefont
  {Rowsell}, \citenamefont {Millward}, \citenamefont {Park},\ and\
  \citenamefont {Yaghi}}]{Rowsell2004}%
  \BibitemOpen
  \bibfield  {author} {\bibinfo {author} {\bibfnamefont {J.~L.~C.}\
  \bibnamefont {Rowsell}}, \bibinfo {author} {\bibfnamefont {A.~R.}\
  \bibnamefont {Millward}}, \bibinfo {author} {\bibfnamefont {K.~S.}\
  \bibnamefont {Park}}, \ and\ \bibinfo {author} {\bibfnamefont {O.~M.}\
  \bibnamefont {Yaghi}},\ }\href {\doibase 10.1021/ja049408c} {\bibfield
  {journal} {\bibinfo  {journal} {J.~Am.~Chem.~Soc.~}\ }\textbf {\bibinfo
  {volume} {126}},\ \bibinfo {pages} {5666} (\bibinfo {year}
  {2004})}\BibitemShut {NoStop}%
\bibitem [{\citenamefont {Yang}\ \emph {et~al.}(2016)\citenamefont {Yang},
  \citenamefont {Fang}, \citenamefont {Ma}, \citenamefont {Pushpa},\ and\
  \citenamefont {Ganz}}]{Yang2016}%
  \BibitemOpen
  \bibfield  {author} {\bibinfo {author} {\bibfnamefont {L.-M.}\ \bibnamefont
  {Yang}}, \bibinfo {author} {\bibfnamefont {G.-Y.}\ \bibnamefont {Fang}},
  \bibinfo {author} {\bibfnamefont {J.}~\bibnamefont {Ma}}, \bibinfo {author}
  {\bibfnamefont {R.}~\bibnamefont {Pushpa}}, \ and\ \bibinfo {author}
  {\bibfnamefont {E.}~\bibnamefont {Ganz}},\ }\href {\doibase
  10.1039/C6CP06981A} {\bibfield  {journal} {\bibinfo  {journal}
  {Phys.~Chem.~Chem.~Phys.~}\ }\textbf {\bibinfo {volume} {18}},\ \bibinfo
  {pages} {32319} (\bibinfo {year} {2016})}\BibitemShut {NoStop}%
\bibitem [{\citenamefont {Blanksby}\ and\ \citenamefont
  {Ellison}(2003)}]{blan+2003acr}%
  \BibitemOpen
  \bibfield  {author} {\bibinfo {author} {\bibfnamefont {S.~J.}\ \bibnamefont
  {Blanksby}}\ and\ \bibinfo {author} {\bibfnamefont {G.~B.}\ \bibnamefont
  {Ellison}},\ }\href {\doibase 10.1021/ar020230d} {\bibfield  {journal}
  {\bibinfo  {journal} {Acc.~Chem.~Res.~}\ }\textbf {\bibinfo {volume} {36}},\
  \bibinfo {pages} {255} (\bibinfo {year} {2003})}\BibitemShut {NoStop}%
\bibitem [{\citenamefont {Edzards}\ \emph {et~al.}(2023)\citenamefont
  {Edzards}, \citenamefont {Sa{\ss}nick}, \citenamefont {Buzanich},
  \citenamefont {Valencia}, \citenamefont {Emmerling}, \citenamefont {Beyer},\
  and\ \citenamefont {Cocchi}}]{edza+2023jpcc}%
  \BibitemOpen
  \bibfield  {author} {\bibinfo {author} {\bibfnamefont {J.}~\bibnamefont
  {Edzards}}, \bibinfo {author} {\bibfnamefont {H.-D.}\ \bibnamefont
  {Sa{\ss}nick}}, \bibinfo {author} {\bibfnamefont {A.~G.}\ \bibnamefont
  {Buzanich}}, \bibinfo {author} {\bibfnamefont {A.~M.}\ \bibnamefont
  {Valencia}}, \bibinfo {author} {\bibfnamefont {F.}~\bibnamefont {Emmerling}},
  \bibinfo {author} {\bibfnamefont {S.}~\bibnamefont {Beyer}}, \ and\ \bibinfo
  {author} {\bibfnamefont {C.}~\bibnamefont {Cocchi}},\ }\href {\doibase
  10.1021/acs.jpcc.3c06054} {\bibfield  {journal} {\bibinfo  {journal}
  {J.~Phys.~Chem.~C}\ }\textbf {\bibinfo {volume} {127}},\ \bibinfo {pages}
  {21456} (\bibinfo {year} {2023})}\BibitemShut {NoStop}%
\bibitem [{\citenamefont {Perdew}\ and\ \citenamefont
  {Levy}(1983)}]{Perdew1983}%
  \BibitemOpen
  \bibfield  {author} {\bibinfo {author} {\bibfnamefont {J.~P.}\ \bibnamefont
  {Perdew}}\ and\ \bibinfo {author} {\bibfnamefont {M.}~\bibnamefont {Levy}},\
  }\href {\doibase 10.1103/PhysRevLett.51.1884} {\bibfield  {journal} {\bibinfo
   {journal} {Phys.~Rev.~Lett.~}\ }\textbf {\bibinfo {volume} {51}},\ \bibinfo
  {pages} {1884} (\bibinfo {year} {1983})}\BibitemShut {NoStop}%
\bibitem [{\citenamefont {Cocchi}\ \emph
  {et~al.}(2011{\natexlab{a}})\citenamefont {Cocchi}, \citenamefont {Ruini},
  \citenamefont {Prezzi}, \citenamefont {Caldas},\ and\ \citenamefont
  {Molinari}}]{cocc+11jpcc}%
  \BibitemOpen
  \bibfield  {author} {\bibinfo {author} {\bibfnamefont {C.}~\bibnamefont
  {Cocchi}}, \bibinfo {author} {\bibfnamefont {A.}~\bibnamefont {Ruini}},
  \bibinfo {author} {\bibfnamefont {D.}~\bibnamefont {Prezzi}}, \bibinfo
  {author} {\bibfnamefont {M.~J.}\ \bibnamefont {Caldas}}, \ and\ \bibinfo
  {author} {\bibfnamefont {E.}~\bibnamefont {Molinari}},\ }\href@noop {}
  {\bibfield  {journal} {\bibinfo  {journal} {J.~Phys.~Chem.~Lett.}\ }\textbf
  {\bibinfo {volume} {115}},\ \bibinfo {pages} {2969} (\bibinfo {year}
  {2011}{\natexlab{a}})}\BibitemShut {NoStop}%
\bibitem [{\citenamefont {Cocchi}\ \emph
  {et~al.}(2011{\natexlab{b}})\citenamefont {Cocchi}, \citenamefont {Prezzi},
  \citenamefont {Ruini}, \citenamefont {Caldas},\ and\ \citenamefont
  {Molinari}}]{cocc+11jpcl}%
  \BibitemOpen
  \bibfield  {author} {\bibinfo {author} {\bibfnamefont {C.}~\bibnamefont
  {Cocchi}}, \bibinfo {author} {\bibfnamefont {D.}~\bibnamefont {Prezzi}},
  \bibinfo {author} {\bibfnamefont {A.}~\bibnamefont {Ruini}}, \bibinfo
  {author} {\bibfnamefont {M.~J.}\ \bibnamefont {Caldas}}, \ and\ \bibinfo
  {author} {\bibfnamefont {E.}~\bibnamefont {Molinari}},\ }\href {\doibase
  10.1021/jz200472a} {\bibfield  {journal} {\bibinfo  {journal}
  {J.~Phys.~Chem.~Lett.}\ }\textbf {\bibinfo {volume} {2}},\ \bibinfo {pages}
  {1315} (\bibinfo {year} {2011}{\natexlab{b}})}\BibitemShut {NoStop}%
\bibitem [{\citenamefont {Krumland}\ \emph {et~al.}(2021)\citenamefont
  {Krumland}, \citenamefont {Valencia},\ and\ \citenamefont
  {Cocchi}}]{krum+21pccp}%
  \BibitemOpen
  \bibfield  {author} {\bibinfo {author} {\bibfnamefont {J.}~\bibnamefont
  {Krumland}}, \bibinfo {author} {\bibfnamefont {A.~M.}\ \bibnamefont
  {Valencia}}, \ and\ \bibinfo {author} {\bibfnamefont {C.}~\bibnamefont
  {Cocchi}},\ }\href@noop {} {\bibfield  {journal} {\bibinfo  {journal}
  {Phys.~Chem.~Chem.~Phys.~}\ }\textbf {\bibinfo {volume} {23}},\ \bibinfo
  {pages} {4841} (\bibinfo {year} {2021})}\BibitemShut {NoStop}%
\end{thebibliography}

\end{document}